\documentclass[12pt]{article}

\usepackage{amssymb,cite}
\usepackage{epsfig}

\makeatletter
\def\eqnumsection{\@addtoreset{equation}{section}
    \def\theequation{\arabic{section}.\arabic{equation}}}
\def\appendixes{\par\setcounter{section}{0}
    \setcounter{subsection}{0}\setcounter{equation}{0}
    \@addtoreset{equation}{section}
    \def\thesection{Appendix \Alph{section}}
    \def\theequation{\Alph{section}.\arabic{equation}}}
\makeatother

\begin{document}

\hyphenation{theo-re-ti-cal}

\begin{flushright}
CERN-TH/2003-063\\
BUTP-2003/04\\
%hep-ph/0303yyy\\
July 2003\\
\end{flushright}

\begin{center}
\centerline{\large\bf The $\eta^\prime g^* g^{(*)}$ Vertex Including the 
$\eta^\prime$-Meson Mass}

\vspace*{1.5cm}

{\large A.~Ali\footnote{On leave of absence from Deutsches
Elektronen-Synchrotron DESY, Hamburg.}
\vskip0.2cm
{\it Theory Division, CERN, CH-122 Geneva 23, Switzerland}}\\
\vspace*{0.3cm}
\centerline{and}
\vspace*{0.3cm}
{\large A.Ya.~Parkhomenko\footnote{On leave of absence from 
Department of Theoretical Physics,
Yaroslavl State University,
Sovietskaya~14, 150000 Yaroslavl, Russia.
}
\vskip0.2cm 
{\it Institut f$\ddot u$r Theoretische Physik, 
Universit$\ddot a$t Bern, \\ 
CH-3012 Bern, Switzerland }}

\vskip0.5cm
{\Large Abstract\\}
\vskip3truemm

\parbox[t]{\textwidth}{
The $\eta^\prime g^* g^{(*)}$ effective vertex function is calculated
in the QCD hard-scattering approach, taking into account
the $\eta^\prime$-meson mass. We work in the approximation in which
only one non-leading Gegenbauer moment for both
the quark-antiquark  and the gluonic light-cone distribution amplitudes 
for the $\eta^\prime$-meson is kept. The vertex function with one 
off-shell gluon is shown to have the form (valid for 
$\vert q_1^2 \vert > m_{\eta^\prime}^2$)  
$F_{\eta^\prime g^* g} (q_1^2, 0, m_{\eta^\prime}^2) = m_{\eta^\prime}^2 
H(q_1^2)/(q_1^2 - m_{\eta^\prime}^2)$, where $H(q_1^2)$ is a slowly 
varying function, derived analytically in this paper. The resulting 
vertex function is in agreement with the phenomenologically inferred 
form of this vertex obtained from an analysis of the CLEO data 
on the $\eta^\prime$-meson energy spectrum in the decay 
$\Upsilon(1S) \to \eta^\prime X$. We also present an interpolating 
formula for the vertex function $F_{\eta^\prime g^* g} (q_1^2, 0, 
m_{\eta^\prime}^2)$ for the space-like region of the virtuality~$q_1^2$, 
which satisfies the QCD anomaly normalization for on-shell gluons 
and the perturbative-QCD result for the gluon virtuality 
$\vert q_1^2\vert \gtrsim 2$ GeV$^2$.
}

\end{center}

\thispagestyle{empty}
\newpage
\setcounter{page}{1}
% Decrease textheight (for preprint numbers) again
\textheight 23.0 true cm

\section{Introduction} 
\label{sec:introduction} 

We reanalyze the $\eta^\prime g^* g^{(*)}$ vertex involving two 
gluons and the $\eta^\prime$-meson, in which one or both of the 
gluons can be virtual. The $\eta^\prime g^* g^{(*)}$ effective 
vertex function (or the $\eta^\prime$~-- gluon transition form factor) 
enters in a number of decays such as $J/\psi \to \eta^\prime \gamma$, 
$B \to (\pi, \rho, K, K^*) \eta^\prime$, $B \to \eta^\prime X_s$,  
$\Upsilon \to \eta^\prime X$, $\Upsilon \to \eta^\prime \gamma$, 
and hadronic production processes, such as 
$N + N(\bar{N}) \to \eta^\prime X$, and hence is of great 
phenomenological importance. Its electromagnetic counterpart, namely 
the vertex $\eta^\prime \gamma^* \gamma$ (equivalently the 
$\eta^\prime - \gamma$ transition form factor) has been measured 
in the process $\gamma \gamma^* \to \eta^\prime$,
and analyzed in the perturbative QCD approach to exclusive processes.
There is no direct experimental measurement of its QCD analogue, 
i.e., the process  $g g^* \to \eta^\prime$, but inclusive decays of 
the heavy mesons such as $B \to \eta^\prime X_s$ and $\Upsilon (1S) 
\to \eta^\prime X$ are probably as close as one could get 
phenomenologically to the underlying vertex.
  
A first attempt to describe the $\eta^\prime g^* g$ vertex in terms 
of the convolution of the distribution amplitudes (DAs) of the 
$\eta^\prime$-meson in the lowest twist (twist-two) approximation and 
a hard-scattering kernel, involving the perturbatively calculable 
processes $\bar q q \to g g$ and $g g \to g g$, was undertaken in 
Ref.~\cite{Muta:1999tc}. In an earlier paper~\cite{Ali:2000ci}, 
we extended this analysis to the case involving two virtual gluons, 
i.e. the vertex $\eta^\prime g^* g^*$, and studied the effect of
including the transverse momentum of the partons in the 
$\eta^\prime$-meson using the Sudakov formalism. Subsequent to this 
detailed study, the $\eta^\prime g^* g^*$ vertex was studied in 
Refs.~\cite{Kroll:2002nt} and~\cite{Agaev:2002ek}, ignoring the 
transverse momentum effects. The latter of the two included power 
corrections in $1/Q^{2n}$ by using the running coupling constant method 
in the standard hard-scattering approach. 
However, both of these papers, as well as the earlier ones, 
neglected the $\eta^\prime$-meson mass effects. 
While ignoring the meson mass is an excellent approximation for 
the pion-photon transition form factor (equivalently the 
$\pi \gamma^* \gamma$ vertex), this is not expected to be quantitatively 
reliable for the case of the $\eta^\prime$-meson due to its large mass,  
$m_{\eta^\prime} = 958$~MeV, in particular for the lower~$Q^2$ 
region, where~$Q^2$ is the virtuality in the $\eta^\prime g^* g^{(*)}$ 
vertex. We present an improved calculation of the $\eta^\prime g^* g^*$ 
effective vertex function including the $\eta^\prime$-meson mass. 
In doing this, we also correct an error in our earlier 
paper~\cite{Ali:2000ci} due to an inappropriate 
choice of the projection operator of the $\eta^\prime$-meson 
onto the two-gluon state.
   
A phenomenological form for the $\eta^\prime - g$ transition 
form factor was proposed by Kagan and Petrov some 
time ago~\cite{Kagan:1997qn}: 
\begin{equation} 
F_{\eta^\prime g^* g} (q_1^2, 0, m_{\eta^\prime}^2) = 
\frac{m_{\eta^\prime}^2\, H (q_1^2, 0, m_{\eta^\prime}^2)}
     {q_1^2 - m_{\eta^\prime}^2} . 
\label{eq:VF-eta-gamma}
\end{equation} 
As an explicit form for the function~$H (q_1^2, 0, m_{\eta^\prime}^2)$, 
these authors suggested to neglect the $q_1^2$-dependence and approximate 
this function as a constant, with $H (q_1^2, 0, m_{\eta^\prime}^2) \simeq 
1.8~{\rm GeV}^{-1}$ extracted from the data on the decay
$J/\psi \to \eta^\prime \gamma$~\cite{Atwood:1997bn}. 
This phenomenological transition form factor is in qualitative 
agreement with the $\eta^\prime$-meson energy 
spectrum near the upper end of the spectrum in the process $\Upsilon (1S) 
\to \eta^\prime X$~\cite{Kagan:2002dq}, 
measured recently by the CLEO collaboration~\cite{Artuso:2002px}, 
and also analyzed by us in a recent paper~\cite{Ali:2003vw}.
We show in this paper that the  
form~(\ref{eq:VF-eta-gamma}) of the $\eta^\prime - g$ transition form 
factor is obtained in the perturbative hard-scattering approach, if a 
light-cone wave-function for the $\eta^\prime$-meson is used 
including the $\eta^\prime$-meson mass.

The validity of the perturbative QCD formalism is expected 
to hold only above a certain gluon virtuality, which for the time-like 
region of~$q_1^2$ must be definitely larger than~$m_{\eta^\prime}^2$
to avoid the pole. For the space-like region, there is no a pole for 
any value of~$q_1^2$ and also the effects of including the transverse 
momentum of the partons in the $\eta^\prime$-meson wave-function are 
numerically not important, as shown in Ref.~\cite{Ali:2000ci}. However, 
also for this case, for low values of the gluon virtuality important 
non-perturbative effects are present and the perturbative QCD formalism 
for the $\eta^\prime g^* g$ vertex is no longer applicable. In our 
numerical calculations, we shall set $Q_0^2 = \mu_0^2 = 2$~GeV$^2$,  
where~$Q_0$ and~$\mu_0$ are the threshold in the evaluation of the 
perturbative QCD kernel and the starting point of the  evolution of 
the $\eta^\prime$-meson distribution amplitudes,  respectively. 
For on-shell gluons, the $\eta^\prime g^* g$ vertex function 
$F_{\eta^\prime g g} (0, 0, m_{\eta^\prime}^2)$ is determined by 
the QCD anomaly which is a non-perturbative result. We propose 
an interpolating formula for the vertex function 
$F_{\eta^\prime g^* g} (q_1^2, 0, m_{\eta^\prime}^2)$, 
which interpolates between the normalization of this function 
for on-shell gluons, as determined from the QCD anomaly, 
and the perturbative QCD hard-scattering result for 
space-like gluon virtualities with $Q^2 \geq Q_0^2$ where 
$Q^2 = |q_1^2| + m_{\eta^\prime}^2$. For the time-like virtuality, 
an interpolating formula for the vertex function remains to be worked 
out, as one must include the transverse-momentum effects, 
which are large at low values of~$q_1^2$, in addition to the
non-perturbative and the $\eta^\prime$-meson mass effects.
  
This paper is organized as follows: In section~\ref{sec:projections}, 
we present the $\eta^\prime$-meson projection operators onto the 
quark-antiquark and the gluonic states, taking into account the 
$\eta^\prime$-meson mass effects, and the 
leading-twist distribution amplitudes for the $\eta^\prime$-meson. 
In section~\ref{sec:epgg-vertex}, we derive the $\eta^\prime g^* g^*$ 
effective vertex function in the perturbative QCD approach 
and discuss the region of its applicability. \
Numerical analysis of the effective vertex function 
is presented in section~\ref{sec:numeric}, where we have used the 
constraints on the Gegenbauer coefficients obtained by us~\cite{Ali:2003vw} 
from the analysis of the data on the $\Upsilon(1S) \to \eta^\prime X$ 
decay, combined with the corresponding constraints from the data on 
the process $\gamma \gamma^* \to \eta^\prime$ presented in 
Ref.~\cite{Kroll:2002nt}. In section~\ref{sec:interpolation}, we 
give an interpolating formula for the $\eta^\prime g^* g$ vertex  
function for the space-like region of the gluon virtuality. 
We conclude with a short summary in section~\ref{sec:conclusion}.

\section{The $\eta^\prime$-Meson Projection Operators}
\label{sec:projections}

\subsection{Projection onto the Quark-Antiquark State}
\label{ssec:quark-projection}

The $\eta^\prime$-meson contains both quark-antiquark and gluonic 
components. Being a pseudoscalar meson, its quark content can be 
described by the matrix element of the bilocal axial-vector operator 
in the SU(3) flavour-singlet state:
\begin{equation} 
{\cal O}^{(q)}_{5 \mu} (x, y) = \frac{1}{\sqrt{N_f}} \, 
\bar \Psi (x) \gamma_\mu \gamma_5 \left [ x, y \right ] \Psi (y) , 
\label{eq:quark-operator} 
\end{equation}
where $\Psi (x) = (u (x), d(x), s(x))$ is the triplet of the light
quark fields in the flavour space and $N_f = 3$. The summation 
over the Dirac, colour and flavour indices is implicitly 
assumed in this bilocal operator. The path-ordered gauge factor,
\begin{equation}
\left [ x, y \right ] = {\cal P} \exp \Bigg \{  
i g_s \int\limits_y^x dz^\mu A^B_\mu (z) \, t_B
\Bigg \} ,
\label{eq:path-order}
\end{equation}
is introduced to ensure gauge invariance of the bilocal 
axial-vector quark operator~(\ref{eq:quark-operator}). In the gauge 
factor~(\ref{eq:path-order}), $g_s$~is the QCD coupling constant, 
$A^B_\mu (z)$ ($B = 1, \ldots, N_c^2 - 1$) is a four-potential of 
the gluonic field, $t_B$~are the generators of the colour $SU (N_c)$ 
group with $N_c = 3$ being the number of the quark colours in QCD, 
and the integration is performed over the straight line connecting 
the points~$y$ and~$x$. 

Taking into account the leading-twist (twist-two), twist-three and 
twist-four  contributions, the matrix element of the 
operator~(\ref{eq:quark-operator}) between the vacuum and the 
$\eta^\prime$-meson states can be presented in the form:   
\begin{equation} 
\big < 0 \big | {\cal O}^{(q)}_{5 \mu} (x, -x) 
\big | \eta^\prime (p) \big > 
= i f_{\eta^\prime}  
\int\limits_0^1 du \, e^{i \xi (p x)} \left \{ p_\mu 
\left [ \phi_{\eta^\prime}^{(q)} (u) + \frac{m_{\eta^\prime}^2}{4} \, 
x^2 \, \mathbb{A} (u) \right ] 
+ \frac{m_{\eta^\prime}^2 x_\mu}{2 (px)} \, \mathbb{B} (u) 
\right \} , 
\label{eq:1} 
\end{equation}
where $p_\mu$ is the four-momentum of the $\eta^\prime$-meson 
($p^2 = m_{\eta^\prime}^2$), $u$ and $1 - u$ are the momentum fractions 
carried by the quark and the antiquark inside the $\eta^\prime$-meson, 
and $\xi = 2u - 1$. The function $\phi_{\eta^\prime}^{(q)} (u)$ is the 
twist-two distribution amplitude (DA), and~$\mathbb{A} (u)$
and~$\mathbb{B} (u)$ contain contributions from operators of twist-two,
twist-three and twist-four. This set of the quark-antiquark two-particle 
DAs is completely analogous to the one defined in Ref.~\cite{Ball:1998je}
for the case of the SU(3) flavour-octet pseudoscalar mesons. 
The decay constant~$f_{\eta^\prime}$ is defined in the local limit 
of the matrix element~(\ref{eq:1}) as
\begin{equation}
\big < 0 \big | {\cal O}^{(q)}_{5 \mu} (0, 0)
\big | \eta^\prime (p) \big > = 
\big < 0 \big |
\frac{1}{\sqrt{N_f}} 
\bar \Psi (0) \gamma_\mu \gamma_5 \Psi (0)
\big | \eta^\prime (p) \big >
= i f_{\eta^\prime} p_\mu .
\label{eq:2}
\end{equation}
Thus, the DAs satisfy the following normalization 
conditions:
\begin{equation}
\int\limits_0^1 du \, \phi_{\eta^\prime}^{(q)} (u) = 1, \qquad 
\int\limits_0^1 du \, \mathbb{B} (u) = 0 . 
\label{eq:3}
\end{equation}
The normalization condition for $\mathbb{A} (u)$ is not fixed by 
Eq.~(\ref{eq:2}). This DA is related to the two- and three-particle 
DAs of lower twists with the help of the equations of motion (see
Eqs.~(3.2) and~(6.11) of Ref.~\cite{Ball:1998je}) and, thus, 
its form (and also the normalization) is implicitly fixed.

We now proceed to work out the DAs of the $\eta^\prime$-meson 
on the light-cone. To that end, we construct 
two light-like vectors~$z_\mu$ and~$P_\mu$ ($z^2 = 0$ 
and~$P^2 = 0$) of the same dimensions as the two vectors~$x_\mu$ 
and~$p_\mu$ at hand: 
\begin{eqnarray} 
z_\mu & = & x_\mu - \frac{p_\mu}{m_{\eta^\prime}^2} 
\left [ (p x) - \sqrt{(p x)^2 - m_{\eta^\prime}^2 x^2} \right ] , 
\label{eq:4} \\
P_\mu & = & p_\mu - \frac{m_{\eta^\prime}^2 x_\mu}
            {(p x) + \sqrt{(p x)^2 - m_{\eta^\prime}^2 x^2}} . 
\label{eq:5}
\end{eqnarray}
Their scalar product has a rather complicated 
form:
\begin{equation} 
(P z) = \frac{2 \left [ (p x)^2 - m_{\eta^\prime}^2 x^2 \right ]}
             {(p x) + \sqrt{(p x)^2 - m_{\eta^\prime}^2 x^2}}~,
\label{eq:6}
\end{equation}
which, however, can be simplified if one assumes that the separation 
between the quark and the antiquark inside the $\eta^\prime$-meson 
is light-like ($x_\mu = z_\mu$ and $x^2 = z^2 = 0$). The light-like 
momentum~$P_\mu$ introduced in Eq.~(\ref{eq:5}) then takes a 
simple form:  
\begin{equation}
P_\mu = p_\mu - \frac{m_{\eta^\prime}^2 z_\mu}{2 (pz)} , \qquad 
(P z) = (p z) = (p x) .     
\label{eq:7}
\end{equation}
In terms of such light-like vectors, one can give   
definitions of the DAs of the $\eta^\prime$-meson on the light-cone 
(similar to the ones for the $\pi$-meson~\cite{Ball:1998je}) 
following from the matrix element of the axial-vector current: 
\begin{equation}
\big < 0 \big | {\cal O}^{(q)}_{5 \mu} (z, -z)
\big | \eta^\prime (p) \big > 
= i f_{\eta^\prime} \int\limits_0^1 du \, e^{i \xi (p z)} 
\left [ P_\mu \, \phi_{\eta^\prime}^{(q)} (u)  
+ \frac{m_{\eta^\prime}^2 z_\mu}{2 (P z)} \, 
g_{\eta^\prime}^{(q)} (u) \right ] .
\label{eq:8} 
\end{equation}
Comparing it with Eq.~(\ref{eq:1}), one obtains the relation: 
\begin{equation}
\mathbb{B} (u) = g_{\eta^\prime}^{(q)} (u) 
               - \phi_{\eta^\prime}^{(q)} (u)~, 
\label{eq:9} 
\end{equation}
implying that the Lorentz invariant amplitudes~$\phi_{\eta^\prime}^{(q)} 
(u)$ and~$\mathbb{B} (u)$ can be interpreted as the $\eta^\prime$-meson 
DAs of the nonlocal axial-vector operator at a strictly light-like 
separation. It is easy to see the usual normalization conditions of the 
light-cone DAs: 
\begin{equation} 
\int\limits_0^1 du \, \phi^{(q)}_{\eta^\prime} (u) = 1, 
\qquad 
\int\limits_0^1 du \, g^{(q)}_{\eta^\prime} (u) = 1. 
\label{eq:LCDAq-normalization}
\end{equation}

Two more bilocal matrix elements define DAs of the twist-three:
\begin{eqnarray} 
\big < 0 \big | 
\frac{1}{\sqrt{N_f}} \bar \Psi (x) \gamma_5 [x, -x] \Psi (-x) 
\big | \eta^\prime (p) \big > & = & 
- i \, f_{\eta^\prime} \, \mu_{\eta^\prime} 
\int\limits_0^1 du \, e^{i \xi (p x)} \phi_p^{(\eta^\prime)} (u) , 
\label{eq:DA-PS} \\ 
\big < 0 \big | \frac{1}{\sqrt{N_f}} 
\bar \Psi (x) \sigma_{\alpha \beta} \gamma_5 [x, -x] \Psi (-x) 
\big | \eta^\prime (p) \big > & = & 
- \frac{i}{3} \, f_{\eta^\prime} \, \mu_{\eta^\prime} 
\left [ 1 - \frac{m_{\eta^\prime}^2}{\mu_{\eta^\prime}^2} \right ] 
\label{eq:DA-PT} \\ 
& \times & (p_\alpha x_\beta - p_\beta x_\alpha) 
\int\limits_0^1 du \, e^{i \xi (p x)} \phi_\sigma^{(\eta^\prime)} (u) ,
\nonumber 
\end{eqnarray}
where $\mu_{\eta^\prime} = m_{\eta^\prime}^2/(2 m_u + 2 m_d + 2 m_s)$ 
is a chirally enhanced factor including the masses~$m_q$ ($q = u, d,
s$) of the light quarks. Note that after the transition to the
light-cone vectors~$P_\mu$ and~$z_\mu$, the forms of Eqs.~(\ref{eq:DA-PS}) 
and~(\ref{eq:DA-PT}) are not changed in the twist-three approximation 
and, thus, the Lorentz invariant twist-three DAs 
coincide with the light-cone twist-three DAs. There  also exist 
twist-three DAs defined by the matrix elements of the three-particle 
(quark-antiquark-gluon) operators. For the light pseudoscalar 
flavour-octet mesons, it has been shown in Ref.~\cite{Ball:1998je}  
that their contributions are numerically small. We expect that this also holds 
for the SU(3) flavour-singlet meson and neglect such contributions.  
In general, one should also take into account the twist-four DAs from the 
matrix elements of three-particle nonlocal operators. The self-consistent set 
of such DAs was introduced in Ref.~\cite{Braun:1988qv, Braun:1989iv} 
for light pseudoscalar flavour-octet mesons; their explicit
updated forms can be found in Ref.~\cite{Ball:1998je}. While the effect 
of the chirally enhanced bilocal matrix elements of 
higher twist operators is an interesting issue to study, we neglect here 
the twist-three and twist-four contributions to the $\eta^\prime$-meson 
light-cone wave-function. 

Starting from Eq.~(\ref{eq:8}), one can calculate the 
light-cone projection operator of the $\eta^\prime$-meson onto the
quark-antiquark state in the leading-twist (twist-two) 
approximation\footnote{The path-ordered gauge factor~(\ref{eq:path-order}) 
is assumed to be between the quark fields and is suppressed to 
simplify the presentation.}: 
\begin{equation} 
\big < 0 \big | 
\bar \Psi_{i \alpha a} (z) \Psi_{j \beta b} (- z) 
\big | \eta^\prime (p) \big > = \frac{i f_{\eta^\prime}}{4 N_c}  
\left [ \gamma_5 (P \gamma) \right ]_{j i} 
\delta_{\beta \alpha} \, \frac{\delta_{b a}}{\sqrt{N_f}} 
\int\limits_0^1 du \, e^{i \xi (p z)} \phi_{\eta^\prime}^{(q)} (u) , 
\label{eq:10} 
\end{equation} 
where the two sets of indices~$(i \alpha a)$ and~$(j \beta b)$ 
describe the components of the quark~$\bar \Psi (z)$ and 
antiquark~$\Psi (- z)$ fields in the Dirac, colour and flavour spaces, 
respectively, and $(P \gamma) = P^\mu \gamma_\mu$. 
After the Fourier transformation of Eq.~(\ref{eq:10}) to the momentum 
space (see Eqs.~(A.4) and~(A.5) in Ref.~\cite{Kroll:2002nt}), 
the result can be presented in the form:
\begin{equation}
\frac{i f_{\eta^\prime}}{4 N_c}   
\left [ \gamma_5 (P \gamma) \right ]_{j i} 
\delta_{\beta \alpha} \, \frac{\delta_{b a}}{\sqrt{N_f}} \, 
\phi_{\eta^\prime}^{(q)} (u) = 
2 (n p) \! \int \frac{d z^-}{2 \pi} \, e^{- i \xi (p z)} 
\big < 0 \big | 
\bar \Psi_{i \alpha a} (z) \Psi_{j \beta b} (- z)
\big | \eta^\prime (p) \big > , 
\label{eq:11}
\end{equation}
where $z_\mu = z^- n_\mu$, and $n_\mu$ is a dimensionless  
light-like vector ($n^2 = 0$). In the momentum space, the quantity 
\begin{equation}
{\cal P}^{(q)}_{j \beta b;i \alpha a} = \frac{1}{4 N_c} 
\left [ \gamma_5 (P \gamma) \right ]_{j i} 
\delta_{\beta \alpha} \, \frac{\delta_{b a}}{\sqrt{N_f}}  
\label{eq:12} 
\end{equation}
can be interpreted as the $\eta^\prime$-meson light-cone projection
operator onto the state of the incoming  quark and 
antiquark~\cite{Terentev:qu}. The quark-antiquark twist-two 
light-cone DA, $\phi_{\eta^\prime}^{(q)} (u)$, can be defined 
as follows:
\begin{equation}
i f_{\eta^\prime} \, \phi_{\eta^\prime}^{(q)} (u) = 
2 \int \frac{d z^-}{2 \pi} \, e^{- i \xi (P z)} 
\big < 0 \big | n^\mu {\cal O}^{(q)}_{5 \mu} (z, -z)
\big | \eta^\prime (p) \big > ~,
\label{eq:13}   
\end{equation}
where the bilocal axial-vector operator is 
defined in Eq.~(\ref{eq:quark-operator}).
Some words about the vector~$n_\mu$ are in order. While well-defined 
in the position space through the light-like separation between the
constituents inside the $\eta^\prime$-meson, this vector is  
arbitrary in the momentum space and its explicit choice is dictated by the 
process at hand. The most natural choice of this vector 
is the four-momentum of some massless particle which appears in a given 
process (see, for example, Ref.~\cite{Kroll:2002nt,Beneke:2002jn}). 
If such a vector is absent, the light-like 
vector~$n_\mu$ should be constructed from existing four-vectors  
explicitly. A representation of the vector $n_\mu$ will be 
given in the next section.

\subsection{Projection onto the Gluonic State} 
\label{ssec:glu-projection}

The gluonic content of the $\eta^\prime$-meson can be described 
with the help of the partially traceless and symmetric bilocal 
gluonic operator\footnote{The method of construction of the 
completely traceless quark-antiquark and gluonic operators can 
be found in Refs.~\cite{Geyer:1999uq} and~\cite{Geyer:2000ig}.}: 
\begin{equation} 
\tilde {\cal O}^{(g)}_{\mu \nu} (x, y) = 
\frac{1}{2} \left [  
G_{\mu \alpha} (x) [x, y] \tilde {G_\nu}^\alpha (y) + 
G_{\nu \alpha} (x) [x, y] \tilde {G_\mu}^\alpha (y) 
\right ] 
- \frac{g_{\mu \nu}}{4} \,
G_{\alpha \beta} (x) [x, y] \tilde G^{\alpha \beta} (y) , 
\label{eq:gluon-operator} 
\end{equation} 
where $G^A_{\mu \nu} (x)$ and $\tilde G^A_{\mu \nu} (x)$ are the 
gluonic field strength tensor and its dual tensor, respectively, 
and the path-ordered gauge factor~$[x, y]$ should be taken here 
in the adjoint representation of the colour $SU (N_c)$ group. 
In the above operator, the summation over the colour indices is 
implicitly assumed.  
Starting from the gluonic operator~(\ref{eq:gluon-operator}), 
one can  introduce the gluonic twist-two, twist-three, and twist-four 
DAs, similar to the quark-antiquark DAs~(\ref{eq:1}) corresponding 
to the axial-vector bilocal operator~(\ref{eq:quark-operator}), 
 as follows: 
\begin{eqnarray} 
\big< 0 \big | \tilde {\cal O}^{(g)}_{\mu \nu} (x, -x)
\big | \eta^\prime (p) \big > & = & 
\frac{f_{\eta^\prime} \, C_F}{2 \sqrt{N_f}}  
\int\limits_0^1 du \, e^{i \xi (p x)} 
\left \{ \left [ 
p_\mu p_\nu - \frac{m_{\eta^\prime}^2}{4} \, g_{\mu \nu} 
\right ] \left [ 
\phi_{\eta^\prime}^{(g)} (u) + m_{\eta^\prime}^2 x^2 \mathbb{C} (u)
\right ] \right. 
\nonumber \\ 
& + & 
\left. \left [ 
2 \, \frac{p_\mu x_\nu + p_\nu x_\mu}{(px)} - g_{\mu \nu} 
\right ] 
\frac{m_{\eta^\prime}^2}{4} \, \mathbb{D} (u) 
\right \} , 
\label{eq:14} 
\end{eqnarray}
where $C_F = (N_c^2 - 1)/(2 N_c)$.

In the local limit, the bilocal gluonic
operator~(\ref{eq:gluon-operator}) vanishes 
[$\tilde {\cal O}^{(g)}_{\mu \nu} (0, 0) = 0$] 
due to the following operator relation: 
\begin{equation} 
G_{\mu \alpha} (0) \tilde {G_\nu}^\alpha (0) = \frac{g_{\mu \nu}}{4} \, 
G_{\alpha \beta} (0) \tilde G^{\alpha \beta} (0) , 
\label{eq:15}
\end{equation}
which gives the normalization conditions for the Lorentz-invariant 
gluonic DAs introduced in Eq.~(\ref{eq:14}): 
\begin{equation}
\int\limits_0^1 du \, \phi_{\eta^\prime}^{(g)} (u) = 0, \qquad
\int\limits_0^1 du \, \mathbb{D} (u) = 0. 
\label{eq:16} 
\end{equation}
These conditions leave an arbitrariness in the choice of a constant 
prefactor in the matrix element~(\ref{eq:14}). The motivation of the 
choice made in this paper will be commented in the next subsection.   
The normalization of the DA $\mathbb{C} (u)$ is not determined 
in the local limit of the operator~(\ref{eq:gluon-operator}), 
but this DA can be related with the other two- and three-particle DAs 
of lower twists using the equations of motion in the same manner 
as has been done for the flavour-octet pseudoscalar 
mesons~\cite{Ball:1998je,Braun:1988qv,Braun:1989iv}. 

Following the discussion in the preceding subsection, let us assume 
that the two gluons are separated by the light-like vector 
$x_\mu = z_\mu$ ($x^2 = z^2 = 0$). 
Thus, the matrix element of the bilocal gluonic
operator can be presented in terms of the gluonic light-cone DAs:
\begin{eqnarray}
\big< 0 \big | \tilde {\cal O}^{(g)}_{\mu \nu} (z, -z) 
\big | \eta^\prime (p) \big > & = &  
\frac{f_{\eta^\prime} C_F}{2 \sqrt{N_f}}  
\int\limits_0^1 du \, e^{i \xi (p z)} \left \{  
P_\mu P_\nu \, \phi_{\eta^\prime}^{(g)} (u) 
\right. 
\label{eq:18} \\ 
& + &  
\left. \left [ 
2 \, \frac{P_\mu z_\nu + P_\nu z_\mu}{(P z)} - g_{\mu \nu} 
\right ] 
\frac{m_{\eta^\prime}^2}{4} \, g_{\eta^\prime}^{(g)} (u)     
\right \} , 
\nonumber  
\end{eqnarray}
where the four-vector~$P_\mu$ is defined in Eq.~(\ref{eq:7}). 
Comparison of this equation with Eq.~(\ref{eq:14}) 
leads to the following relation: 
\begin{equation} 
\mathbb{D} (u) = g_{\eta^\prime}^{(g)} (u) 
               - \phi_{\eta^\prime}^{(g)} (u) . 
\label{eq:19} 
\end{equation}
This shows that the Lorentz-invariant DAs, $\phi_{\eta^\prime}^{(g)}
(u)$ and~$\mathbb{D} (u)$, are connected with the light-cone DAs of 
the $\eta^\prime$-meson defined by the bilocal 
operator~(\ref{eq:gluon-operator}) with a strictly light-like 
separation between the gluon fields, in the same manner as for the 
quark-antiquark DAs, $\phi_{\eta^\prime}^{(q)} (u)$ 
and~$\mathbb{B} (u)$, defined by the bilocal axial-vector  
operator~(\ref{eq:quark-operator}). 
The normalization conditions for the gluonic light-cone DAs are: 
\begin{equation}
\int\limits_0^1 du \, \phi_{\eta^\prime}^{(g)} (u) = 0 , \qquad 
\int\limits_0^1 du \, g_{\eta^\prime}^{(g)} (u) = 0 . 
\label{eq:20}
\end{equation}  
As we restricted ourselves to the leading-twist part of the 
quark-antiquark component of the $\eta^\prime$-meson only, 
we also do not consider any further the higher-twist DAs in the  
gluonic component of the $\eta^\prime$-meson.  

The gauge-invariant definition of the gluonic light-cone 
DAs~(\ref{eq:18}) is given in terms of the gluonic field 
strength tensor~$G^A_{\mu \nu} (z)$ and its dual~$\tilde G^A_{\mu \nu}(-z)$,
though the usual Feynman rules involve the gluonic 
four-potential~$A^A_\mu (z)$. A possible way 
to get the required matrix element is to use the relation between 
the field strength tensor and the four-potential in the light-cone 
gauge ($n^\mu A^A_\mu (x; n) = 0$)~\cite{Radyushkin:1996ru}: 
\begin{equation} 
A^A_\mu (x; n) = n^\nu \int\limits_0^\infty 
G^A_{\mu \nu} (x + \sigma n) \, d\sigma , 
\label{eq:A-G-LC-relation}
\end{equation} 
so that it can be applied to Eq.~(\ref{eq:18}) after its contraction 
with the light-like vectors~$n^\mu$ and~$n^\nu$. 
In the leading-twist approximation, the projection operator of the 
$\eta^\prime$-meson onto the gluonic state in terms of the gluonic 
four-potentials can be written as follows~\cite{Kroll:2002nt}\footnote{
Starting from here we suppress the path-ordered gauge factor $[z, -z]$
in the matrix elements.}: 
\begin{equation} 
\big < 0 \big | A^A_{[ \mu} (z) A^B_{\nu ]} (-z) 
\big | \eta^\prime (p) \big > = 
\frac{f_{\eta^\prime} C_F}{4 \sqrt{N_f}} \,  
\frac{\delta_{A B}}{2 N_c C_F} \, 
\frac{\varepsilon_{\mu \nu \rho \sigma} z^\rho P^\sigma}{(z P)} 
\int\limits_0^1 du \, e^{i \xi (p z)} \, 
\frac{\phi_{\eta^\prime}^{(g)} (u)}{u (1 - u)} ,   
\label{eq:21} 
\end{equation}
where $A^A_{[\mu} (x) A^B_{\nu]} (y) \equiv 
[A^A_\mu (x) A^B_\nu (y) - A^A_\nu (x) A^B_\mu (y)]/2$ is the 
bilocal operator antisymmetrized in the Lorentz indices. 
The gluonic matrix element~(\ref{eq:21}) contains the Lorentz structure 
$\varepsilon_{\mu \nu \rho \sigma} \, z^\rho P^\sigma /(z P)$.
To perform the Fourier transform to the momentum space, it is convenient
to introduce the dimensionless light-like vector~$n_\alpha$ (its explicit
form in a specific frame is given in Eq.~(A.3) of Ref.~\cite{Kroll:2002nt}) 
so that $z_\alpha = z^- n_\alpha$. The light-like vector~$P_\mu$ defined
by Eq.~(\ref{eq:7}) becomes independent of the variable~$z^-$:
\begin{equation}
P_\mu = p_\mu - \frac{m_{\eta^\prime}^2 n_\mu}{2 (p n)}, \qquad   
(P n) = (p n),
\label{eq:22}
\end{equation}
and the Lorentz structure under consideration transforms as
\begin{equation}
\varepsilon_{\mu \nu \rho \sigma} \, \frac{z^\rho P^\sigma}{(z P)}\, 
\longrightarrow \,
\varepsilon_{\mu \nu \rho \sigma} \, \frac{n^\rho P^\sigma}{(n P)} = 
\varepsilon_{\mu \nu \rho \sigma} \, \frac{n^\rho p^\sigma}{(n p)} .
\label{eq:23} 
\end{equation}
Note the invariance of this Lorentz structure under the
transformations $P_\mu \to P_\mu^\prime = P_\mu + C \, n_\mu$ or 
$n_\mu \to n_\mu^\prime = n_\mu + \tilde C \, P_\mu$, where~$C$ 
and~$\tilde C$ are arbitrary factors. The last expression in 
Eq.~(\ref{eq:23}) corresponds to the specific choice of the first 
factor: $C = m_{\eta^\prime}^2/2 (np)$. The invariance of the 
Lorentz structure~(\ref{eq:23}) under the transformation of the 
light-like vector~$n_\mu$ was pointed out in Ref.~\cite{Kroll:2002nt} 
for the case of the massless $\eta^\prime$-meson 
(i.e., with $p^2 = 0$ and $P_\mu = p_\mu$).

After performing the Fourier transformation of Eq.~(\ref{eq:21}), 
we obtain the following result:
\begin{equation}
\frac{f_{\eta^\prime} \, \delta_{A B}}{8 N_c \sqrt{N_f}} \, 
\frac{\varepsilon_{\mu \nu \rho \sigma} n^\rho p^\sigma}{(n p)} \, 
\frac{\phi_{\eta^\prime}^{(g)} (u)}{u (1 - u)} 
= 2 (n p) \int \frac{d z^-}{2 \pi} \, e^{- i \xi (p z)}
\big < 0 \big | A^A_{[\mu} (z) A^B_{\nu]} (-z) 
\big | \eta^\prime (p) \big > . 
\label{eq:24}
\end{equation}
In the momentum space, the quantity  
\begin{equation}
{\cal P}^{(g)}_{\mu A; \nu B} =   
\frac{i \, \delta_{A B}}{4 N_c \sqrt{N_f}} \, 
\frac{\varepsilon_{\mu \nu \rho \sigma} n^\rho p^\sigma}{(n p)}  
\label{eq:25}
\end{equation}
is the projection operator of the $\eta^\prime$-meson onto the state 
of two incoming gluons~\cite{Terentev:qu}. The gluonic twist-two 
light-cone DA can be defined through the following matrix element:
\begin{equation}
f_{\eta^\prime} \, \phi_{\eta^\prime}^{(g)} (u) = 
\frac{4 \sqrt{N_f}}{C_F (n p)} 
\int \frac{d z^-}{2 \pi} \, e^{- i \xi (p z)} \big < 0 \big | 
n^\mu G_{\mu \alpha} (z) n_\nu \tilde G^{\nu \alpha} (-z) 
\big | \eta^\prime (p) \big > ,
\label{eq:26}   
\end{equation}
where the summation over the colour indices is implicitly assumed 
on the right-hand side.

\subsection{The $\eta^\prime$-Meson Light-Cone Distribution Amplitudes} 
\label{ssec:DAs}

 The leading-twist light-cone DAs of the 
$\eta^\prime$-meson can be presented as infinite series in the 
Gegenbauer polynomials $C^{3/2}_n (u - \bar u)$ for the quark-antiquark 
component and $C^{5/2}_{n -1} (u - \bar u)$ for the gluonic 
one~\cite{Ohrndorf:1981uz,Shifman:1981dk,Baier:1981pm,Terentev:wv,Terentev:qu,Belitsky:1998gc}:  
\begin{eqnarray}
\phi^{(q)}_{\eta^\prime} (u, Q^2) & = & 6 u \bar u
\left [ 1 +  
\sum_{{\rm even} \, n \ge 2} A_n (Q^2) \, C^{3/2}_n (u - \bar u)
\right ] , 
\label{eq:DAq-gen} \\  
\phi^{(g)}_{\eta^\prime} (u, Q^2) & = & u^2 \bar u^2 
\sum_{{\rm even} \, n \ge 2} 
B_n (Q^2) \, C^{5/2}_{n - 1} (u - \bar u) , 
\label{eq:DAg-gen}
\end{eqnarray}
where $\bar u = 1 - u$, and the following notations are introduced 
for the Gegenbauer moments:
\begin{eqnarray} 
A_n (Q^2) & = & B^{(q)}_n (\mu_0^2) \left [ 
\frac{\alpha_s (\mu_0^2)}{\alpha_s (Q^2)} \right ]^{\gamma_+^n}
+ \rho_n^{(g)} \, B^{(g)}_n (\mu_0^2) \left [ 
\frac{\alpha_s (\mu_0^2)}{\alpha_s (Q^2)} \right ]^{\gamma_-^n} , 
\label{eq:An} \\
B_n (Q^2) &= & \rho_n^{(q)} \, B^{(q)}_n (\mu_0^2) \left [ 
\frac{\alpha_s (\mu_0^2)}{\alpha_s (Q^2)} \right ]^{\gamma_+^n}
+ B^{(g)}_n (\mu_0^2) \left [ 
\frac{\alpha_s (\mu_0^2)}{\alpha_s (Q^2)} \right ]^{\gamma_-^n} . 
\label{eq:Bn} 
\end{eqnarray} 
These equations contain the quantities~$\gamma_{\pm}^n$, defined 
as:
\begin{equation}
\gamma_{\pm}^n \equiv  \frac{1}{2} \,
\left [ \gamma_{QQ}^n + \gamma_{GG}^n
\pm \sqrt{(\gamma_{QQ}^n - \gamma_{GG}^n)^2 +
          4 \gamma_{QG}^n \gamma_{GQ}^n} \,
\right ],
\label{eq:AD-rotate}
\end{equation}
where the anomalous dimensions~$\gamma_{ij}^n$ ($i,j = Q,G$) are
%~\cite{Belitsky:1998gc}:
%
\begin{eqnarray}
\gamma_{QQ}^n & = & \frac{C_F}{\beta_0}
\left [ 3 + \frac{2}{(n + 1)(n + 2)} -
        4 \sum_{j = 1}^{n + 1} \frac{1}{j} \right ] ,
\nonumber \\
\gamma_{QG}^n & = & \frac{C_F}{\beta_0} \,
\frac{n (n + 3)}{3 (n + 1) (n + 2)} ,
\label{eq:AD-matrix} \\
\gamma_{GQ}^n & = & \frac{N_f}{\beta_0} \,
\frac{12}{(n + 1)(n + 2)} ,
\nonumber \\
\gamma_{GG}^n & = & \frac{N_c}{\beta_0}
\left [ \frac{8}{(n + 1)(n + 2)} -
        4 \sum_{j = 1}^{n + 1} \frac{1}{j} \right ] + 1 .  
\nonumber
\end{eqnarray}
Here, $\beta_0 = (11 N_c - 2 n_f)/3$ is the first coefficient in 
the expansion of the QCD $\beta$-function and~$n_f$ is the number 
of quarks with masses less than the energy scale~$Q$ entering 
in the DAs~(\ref{eq:DAq-gen}) and~(\ref{eq:DAg-gen}). Note the  
difference between the quark numbers~$N_f = 3$, used earlier in the 
context of the wave-function of the $\eta^\prime$-meson, and~$n_f$ 
in the $\beta$-function; the former reflects the quark content 
of the SU(3) flavour-singlet meson, while the latter is the number of  
the active quarks, which (together with the gluons) determine the 
renormalization effects in the LCDAs of the $\eta^\prime$-meson, 
probed at a virtuality~$Q^2$.
The discussion of these anomalous dimensions in the one- 
and two-loop accuracy can be found in Ref.~\cite{Belitsky:1998gc}.
Our choice of the constant prefactor in the gluonic 
matrix element~(\ref{eq:14}) is reflected in the non-diagonal anomalous 
dimensions presented above.

The anomalous dimensions~(\ref{eq:AD-matrix}) also define   
the quantities~$\rho^{(g)}_n$ and~$\rho^{(q)}_n$ entering in the 
Gegenbauer moments~(\ref{eq:An}) and~(\ref{eq:Bn}) as follows:
\begin{equation}
\rho^{(q)}_n = 6 \, \frac{\gamma_+^n - \gamma_{QQ}^n}{\gamma_{QG}^n} ,
\qquad
\rho^{(g)}_n = \frac{1}{6} \,
\frac{\gamma_{QG}^n}{\gamma_-^n - \gamma_{QQ}^n} .
\label{eq:rho's}
\end{equation}
The prefactor in the gluonic light-cone DA~(\ref{eq:DAg-gen}) 
is chosen as $u^2 \bar u^2$ according to Ref.~\cite{Kroll:2002nt}, 
which accounts for the appearance of the factor~$u \bar u$ 
in the denominator of the integral function in the gluonic matrix 
element~(\ref{eq:21}). Such a functional dependence is also in agreement 
with the general expression for the ``asymptotic'' form of the LCDAs 
resulting from the conformal symmetry~\cite{Ball:1998je,Braun:1989iv}. 

We use an approximate form for the $\eta^\prime$-meson light-cone 
wave-function in which only the first non-asymptotic term in 
both the quark-antiquark and gluonic DAs is kept. Thus,  
\begin{equation} 
\phi^{(q)}_{\eta^\prime} (u, Q^2) = 6 u \bar u  
\left [ 1 + 6 (1 - 5 u \bar u) \, A_2 (Q^2) \right ] ,  
\qquad 
\phi^{(g)}_{\eta^\prime} (u, Q^2) = 
5 u^2 \bar u^2 \, (u - \bar u) \, B_2 (Q^2) , 
\label{eq:DAqg}
\end{equation}
where the explicit forms of the Gegenbauer polynomials 
$C^{3/2}_2 (u - \bar u)$ and $C^{5/2}_1 (u - \bar u)$ have been used. 
The free parameters~$B^{(q)}_2 (\mu_0^2)$ and~$B^{(g)}_2 (\mu_0^2)$ 
(the Gegenbauer coefficients), which enter in the Gegenbauer 
moments~$A_2(Q^2)$ and~$B_2(Q^2)$, are not determined from first 
principles and have to be modeled or extracted from a phenomenological 
analysis of experimental data. We defer the quantitative discussion 
of these Gegenbauer coefficients to Sec.~\ref{sec:numeric}. 

There is an additional point concerning the scale~$\mu_0^2$. The 
usual choice of the scale $\mu_0^2 = 1$~GeV$^2$ made in the analysis 
of the pion form factor and the $\pi^0 - \gamma$ transition form 
factor represents the initial value for the evolution of the 
LCDAs. As the mass of the $\eta^\prime$-meson is of order 1~GeV, a more 
realistic choice of the parameter $\mu_0^2$ is $\mu_0^2 =2$~GeV$^2$, 
which we shall adopt. Consistent with this assumption, and with the 
choice $Q^2 = |q^2| + m_{\eta^\prime}^2$, where $q^2$ is the total 
gluon virtuality in the $\eta^\prime g^* g^*$ vertex, we shall also set 
$Q_0^2 = 2$~GeV$^2$ in the calculation of the perturbative kernel.

\section{The $\eta^\prime g^* g^*$ Effective Vertex Function} 
\label{sec:epgg-vertex}

In the momentum space, the effective $\eta^\prime g^* g^*$ vertex 
can be extracted from the invariant matrix element of the process 
$\eta^\prime \to g^* g^*$ with the help of the relation\footnote{
The difference in the phase factor~$i$ between the definition of the 
effective $\eta^\prime g^* g^*$ vertex function in this  paper 
and in Ref.~\cite{Ali:2000ci} is related to the corresponding 
difference in the definitions of the projection operators of the 
$\eta^\prime$-meson onto the quark-antiquark and gluonic states.}: 
\begin{equation}
{\cal M} = {\cal M}^{(q)} + {\cal M}^{(g)} \equiv 
F_{\eta^\prime g^* g^*} (q_1^2, q_2^2, m_{\eta^\prime}^2) \,
\delta_{A B} \, \varepsilon^{\mu \nu \rho \sigma} \,
\varepsilon^{A*}_{1\mu} \varepsilon^{B*}_{2\nu} q_{1\rho} q_{2\sigma} ,
\label{eq:EV-def}
\end{equation}
where ${\cal M}^{(q)}$ and ${\cal M}^{(g)}$ are the contributions 
from the quark-antiquark and gluonic components of the
$\eta^\prime$-meson, respectively, and $q_i$ and $\varepsilon^A_i$ 
($i = 1, 2$) are the four-momenta and the polarization vectors of the 
final virtual gluons. The four-momentum of the $\eta^\prime$-meson is 
related to the four-momenta of the gluons by energy-momentum 
conservation: $p_\mu = q_{1 \mu} + q_{2 \mu}$.   
The individual contributions~${\cal M}^{(q)}$ and~${\cal M}^{(g)}$ to 
the invariant amplitude~(\ref{eq:EV-def}) can be calculated by using 
the $\eta^\prime$-meson projection operators onto the 
quark-antiquark~(\ref{eq:12}) and the two-gluonic~(\ref{eq:25}) states,
yielding 
\begin{eqnarray}
{\cal M}^{(q)} & = & i f_{\eta^\prime} 
\int\limits_0^1 du \, \phi^{(q)}_{\eta^\prime} (u, Q^2) \, 
{\cal P}^{(q)}_{j \beta b; i \alpha a} \, 
\delta^{a b} \big [ T^{(q)}_{\rm H} \big ]^{\alpha \beta}_{i j} , 
\label{eq:MEqq-def} \\
{\cal M}^{(g)} & = & 
\frac{i f_{\eta^\prime}}{2} \int\limits_0^1 du \, 
\frac{\phi^{(g)}_{\eta^\prime} (u, Q^2)}{u \bar u} \,
{\cal P}^{(g)}_{\sigma D; \rho C} \,
\big [ T^{(g)}_{\rm H} \big ]^{\rho \sigma}_{C D} ,
\label{eq:MEgg-def}  
\end{eqnarray}
where $T_{\rm H}^{(q)}$ and $T_{\rm H}^{(g)}$ are the quark-antiquark and 
gluonic hard-scattering kernels calculated in the perturbative QCD, 
respectively.  
The factor~$1/2$ in Eq.~(\ref{eq:MEgg-def}) takes 
into account the two identical gluons in the $\eta^\prime$-meson. 
To go further, it is necessary to 
define the light-like vector~$n_\mu$ appearing in the individual 
invariant amplitudes~(\ref{eq:MEqq-def}) and~(\ref{eq:MEgg-def}) in 
terms of the physical vectors of the problem under study.

\subsection{Specifying the Light-Like Vector $n_\mu$}
\label{ssec:n-mu}   

We now proceed to express the vector~$n_\mu$ in terms of the
gluon momenta~$q_{1\mu}$ and~$q_{2\mu}$ in the $\eta^\prime g^* g^*$ 
vertex. In particular, we 
consider the case when both the gluons are off the mass shell and have 
virtualities~$q_1^2$ and~$q_2^2$ comparable to the 
$\eta^\prime$-meson mass squared.

Let us assume, to be definite, that the virtualities of both gluons 
are time-like ($q_1^2 > 0$ and $q_2^2 > 0$). In that case, the 
two light-like vectors~$n^{(+)}_\mu$ and~$n^{(-)}_\mu$ can 
be constructed from~$q_{1\mu}$
and~$q_{2\mu}$ (or, equivalently, from~$p_\mu = q_{1\mu} + q_{2\mu}$
and~$q_{2\mu}$):   
\begin{equation}
n^{(\pm)}_\mu = C_n \left \{
2 q_2^2 p_\mu + \left [ q_1^2 - q_2^2 - m_{\eta^\prime}^2 \pm
\lambda_{\rm K} \Big ( m_{\eta^\prime}, \sqrt{q_1^2}, \sqrt{q_2^2} \Big )     
\right ] q_{2 \mu}
\right \} ,
\label{eq:n-pm-gen}
\end{equation}
where $C_n$ is an arbitrary factor and 
$\lambda_{\rm K} (a, b, c)$ is a kinematic function, defined as follows:
\begin{equation}
\lambda_{\rm K} (a, b, c) = \sqrt{ \left ( a + b + c \right )
\left ( a - b - c \right ) \left ( a - b + c \right )     
\left ( a + b - c \right ) } .
\label{eq:lambda-function}
\end{equation}
In the limit of neglecting the $\eta^\prime$-meson mass  
(i.e., setting $m_{\eta^\prime} = 0$), these vectors are 
simplified to:
\begin{equation}
n^{(\pm)}_\mu = C_n \left \{
2 q_2^2 p_\mu + \left [ q_1^2 - q_2^2 \pm | q_1^2 - q_2^2 | \right ]  
q_{2\mu} \right \} ,
\label{eq:n-pm-0}
\end{equation}
as $\lambda_{\rm K} (0, \sqrt{q_1^2}, \sqrt{q_2^2}) = | q_1^2 - q_2^2 |$.
Thus, if $q_1^2 > q_2^2$, then these vectors have the forms:
\begin{eqnarray}
n^{(+)}_\mu & = & 2 C_n
\left ( q_2^2 q_{1 \mu} + q_1^2 q_{2 \mu} \right ) ,
\label{eq:n-pm->} \\
n^{(-)}_\mu & = & 2 C_n \,  q_2^2 p_\mu ,
\nonumber
\end{eqnarray}
while for $q_1^2 < q_2^2$ they are:
\begin{eqnarray}
n^{(-)}_\mu & = & 2 C_n
\left ( q_2^2 q_{1 \mu} + q_1^2 q_{2 \mu} \right ) ,
\label{eq:n-pm-<} \\
n^{(+)}_\mu & = & 2 C_n \,  q_2^2 p_\mu . 
\nonumber
\end{eqnarray}
If we take $n_\mu = n^{(+)}_\mu$ ($n_\mu = n^{(-)}_\mu$), then in the 
limit $m_{\eta^\prime} = 0$, the light-like vector $P_\mu$~(\ref{eq:22}) 
vanishes for $q_1^2 < q_2^2$ ($q_1^2 > q_2^2$) and, hence, the
projection operators onto the quark-antiquark~(\ref{eq:12}) and 
gluonic~(\ref{eq:25}) states also vanish. To get non-vanishing 
projection operators in both regions, the vector~$n_\mu$ should be 
taken as:
\begin{equation}
n_\mu = n^{(+)}_\mu \, \Theta (q_1^2 - q_2^2) +
        n^{(-)}_\mu \, \Theta (q_2^2 - q_1^2) ,
\label{eq:n-pm-<>}
\end{equation}
where $\Theta (x)$ is the unit step function.
To write this vector uniformly, it is convenient to 
introduce the total gluon virtuality~$q^2$, the asymmetry 
parameter~$\omega$, and the relative $\eta^\prime$-meson mass 
squared~$\eta$ as follows: 
\begin{equation} 
q^2 = q_1^2 + q_2^2, \qquad 
\omega = \frac{q_1^2 - q_2^2}{q^2}, \qquad 
\eta = \frac{m_{\eta^\prime}^2}{q^2} . 
\label{eq:parameters}
\end{equation} 
If we also redefine the factor $C_n$ as $C_n = \tilde C_n/q^2$, the 
vector $n_\mu$~(\ref{eq:n-pm-<>}) can be rewritten in the form:
\begin{equation}
n_\mu = \tilde C_n \left [ (1 - \omega) \, p_\mu +
        (\omega - \eta + \omega \lambda) \, q_{2 \mu} \right ] ,
\label{eq:n-res}
\end{equation}
where the function $\lambda$ can be written as\footnote{Note, 
$\lambda_{\rm K}$ and $\lambda$ are different functions.}:
\begin{equation}
\lambda = 
\sqrt{1 - \frac{2 \eta}{\omega^2} + \frac{\eta^2}{\omega^2}} .
\label{eq:lambda}
\end{equation}
In the massless limit of the $\eta^\prime$-meson ($\eta = 0$),  
$\lambda=1$.
With this specification of the light-like vector~$n_\mu$, its scalar 
product with the $\eta^\prime$-meson four-momentum~$p_\mu$ is:
\begin{equation}
(p n) = - \frac{q^2}{2} \, \tilde C_n \omega \lambda
\left ( \omega - \eta + \omega \lambda \right ) ,
\label{eq:pn-scalar}
\end{equation}
which allows us to write the light-like vector~$P_\mu$~(\ref{eq:22})
in the following form:
\begin{equation}
P_\mu = \frac{1}{2 \omega \lambda}
\left [ (\omega - \eta + \omega \lambda) \, p_\mu
       + 2 \eta \, q_{2 \mu} \right ] .
\label{eq:P-res}
\end{equation}
With this, the Lorentz structure in the projection 
operator~(\ref{eq:25}) is reduced to the form:
\begin{equation} 
\frac{\varepsilon_{\mu \nu \rho \sigma} n^\rho p^\sigma}{(n p)} = 
\frac{2}{\omega \lambda} \, 
\frac{\varepsilon_{\mu \nu \rho \sigma} q_1^\rho q_2^\sigma}{q^2} =
\frac{2}{\lambda} \,
\frac{\varepsilon_{\mu \nu \rho \sigma} q_1^\rho q_2^\sigma}
     {q_1^2 - q_2^2} .
\label{eq:glu-PO-real}
\end{equation} 
Note that both the vector $P_\mu$, defining the 
Dirac structure of the $\eta^\prime$-meson projection operator 
onto the quark-antiquark state~(\ref{eq:12}), and the Lorentz 
structure considered above, which comes from the projection operator
onto the gluonic state~(\ref{eq:12}), are independent of the 
choice of the factor~$\tilde C_n$. 

For the case when both the gluons have space-like virtualities 
($q_1^2 < 0$ and $q_2^2 < 0$), the same set of quantities can be 
introduced as in Eq.~(\ref{eq:parameters}), with the 
obvious difference that the total gluon virtuality and the 
relative $\eta^\prime$-meson mass squared are negative. This 
difference does not change the result obtained for the 
time-like gluon virtualities and, thus, Eqs.~(\ref{eq:n-res}), 
(\ref{eq:P-res}) and~(\ref{eq:glu-PO-real}) are valid in this 
case also.

\subsection{The Quark Part of the $\eta^\prime g^* g^*$ Vertex} 
\label{ssec:epgg-quark-result}

\begin{figure}[t]
\centerline{
\psfig{file=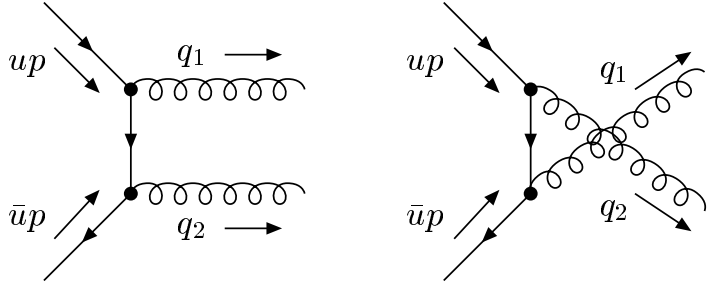,width=.65\textwidth} 
}
\caption{\label{fig:diag-qu}% 
         Leading Feynman diagrams contributing to the quark  
         part of the $\eta^\prime g^* g^*$ vertex.} 
\end{figure}

The light-like vector~$P_\mu$, and hence the $\eta^\prime$-meson
projection operator~(\ref{eq:12}) onto the quark-antiquark state, is 
completely defined in terms of the physical vectors~-- the four-momenta 
of the gluons. Hence, we can calculate the quark part of the 
$\eta^\prime g^* g^*$ effective vertex function starting from the 
invariant amplitude~(\ref{eq:MEqq-def}), obtaining the hard-scattering 
kernel: 
\begin{equation}
\big [ T^{(q)}_{\rm H} \big ]^{\alpha \beta} = 
V^{\alpha \beta; A B}_{\mu \nu} (u p, \bar u p, - q_1, - q_2) \, 
\varepsilon^{A*}_{1\mu} \varepsilon^{B*}_{2\nu} , 
\label{eq:Tq-def}
\end{equation}
where the Dirac indices~$i$ and~$j$ are not shown explicitly; 
the exact expression for the effective quark-antiquark-gluon-gluon 
vertex in the lowest order in perturbative QCD can be found in 
Eq.~(3.3) of Ref.~\cite{Ali:2000ci}. This yields the following 
result: 
\begin{equation}
{\cal M}^{(q)} = - \frac{i f_{\eta^\prime} \sqrt{N_f}}{4 N_c} \, 
4 \pi \alpha_s (Q^2) \, \delta_{A B} \!\!
\int\limits_0^1 du \, \phi^{(q)}_{\eta^\prime} (u, Q^2) \, 
{\rm Sp} \left \{ \gamma_5 (P \gamma) 
\frac{(\varepsilon^{A*}_1 \gamma) ([u p - q_2] \gamma)  
      (\varepsilon^{B*}_2 \gamma)}{(u p - q_2)^2 + i \epsilon} 
\right \} , 
\label{eq:MEq-explicitl}
\end{equation}  
where $Q^2 = |q^2| + m_{\eta^\prime}^2$ and the summation over the 
quark colour and flavour indices have been performed. 
This matrix element contains the 
contributions from both diagrams presented in Fig.~\ref{fig:diag-qu}, 
as the contribution from the second diagram can be transformed to the 
form of the first one with the help of the symmetry property of the 
quark-antiquark light-cone DA: $\phi^{(q)}_{\eta^\prime} (u, Q^2) = 
\phi^{(q)}_{\eta^\prime} (\bar u, Q^2)$. 
It is easy to see that the parameter~$C$ defined by Eq.~(2.3) of 
Ref.~\cite{Ali:2000ci} is connected with the $\eta^\prime$-meson decay 
constant~$f_{\eta^\prime}$ used here by the following relation: 
$C = \sqrt{N_f} \, f_{\eta^\prime}$.  
%
%%%%%%%%%%%%%%%%%%%%%%%%%%%%%%%%%%%%%%%%%%%%%%%%%%%%%%%%%%%%%%%%%%%%%%%%%
%
%\hspace{90mm}
%
%%%%%%%%%%%%%%%%%%%%%%%%%%%%%%%%%%%%%%%%%%%%%%%%%%%%%%%%%%%%%%%%%%%%%%%%%
%

Taking all this into account as well as the Ansatz~(\ref{eq:EV-def}) 
for extracting the $\eta^\prime g^* g^*$ effective vertex function, 
the result for the quark part can be written as follows: 
\begin{equation}
F^{(q)}_{\eta^\prime g^* g^*} (q^2, \omega, \eta)
= \frac{4 \pi \alpha_s (Q^2)}{q^2} \, 
\frac{f_{\eta^\prime} \sqrt{N_f}}{N_c \, \omega \lambda}
\int\limits_0^1 du \, \phi^{(q)}_{\eta^\prime} (u, Q^2) \,
\frac{\omega (1 + \lambda) + \eta (u - \bar u)}
     {1 + \omega (u - \bar u) - 2 u \bar u \eta + i \epsilon} . 
\label{eq:epgg-q0}
\end{equation}
The quark-antiquark contribution, in the approximation of keeping the 
first two terms in the corresponding distribution amplitude 
$\phi^{(q)}_{\eta^\prime} (u, Q^2)$ [see Eq.~(\ref{eq:DAqg})], 
can be conveniently written in the following form 
(similar to Eq.~(3.7) of Ref.~\cite{Ali:2000ci}):
\begin{equation}
F^{(q)}_{\eta^\prime g^* g^*} (q^2, \omega, \eta) =
\frac{4 \pi \alpha_s (Q^2)}{m_{\eta^\prime}^2 \, \lambda} \, 
\frac{3 f_{\eta^\prime} \sqrt{N_f}}{N_c}
\left \{ G^{(q)}_0 (\omega, \eta) + 6 A_2 (Q^2) G^{(q)}_2 (\omega, \eta)
\right \},
\label{eq:epgg-res-gen}
\end{equation}
where $A_2 (Q^2)$ is the Gegenbauer moment defined by Eq.~(\ref{eq:An}),  
and the functions~$G^{(q)}_0 (\omega, \eta)$
and~$G^{(q)}_2 (\omega, \eta)$ are:
\begin{eqnarray}
G^{(q)}_0 (\omega, \eta) & = & 
1 - \lambda +
\left [ 1 - \frac{\omega^2}{\eta} (1 - \lambda) \right ]
\left [ \frac{1}{2 \omega} \ln
\left | \frac{1 + \omega}{1 - \omega} \right | +
\lambda \, {\rm J} (\omega, \eta) \right ] , 
\label{eq:Gq0} \\
G^{(q)}_2 (\omega, \eta) & = & \frac{5}{2 \eta}
\left \{ \frac{\omega^2}{\eta} (1 - \lambda)^2 -
\left ( 1 - \frac{\eta}{15} \right ) (1 - \lambda)
\right.
\label{eq:Gq2} \\
& - & \left.
\left [ 1 - \frac{\omega^2}{\eta} (1 - \lambda) \right ]
\left [
1 - \frac{2 \eta}{5} - \frac{\omega^2}{\eta} (1 - \lambda)
\right ]
\left [ \frac{1}{2 \omega} \ln
\left | \frac{1 + \omega}{1 - \omega} \right | +
\lambda \, {\rm J} (\omega, \eta) \right ]
\right \} . 
\nonumber
\end{eqnarray}
The results given above are presented using the function:
\begin{equation}
{\rm J} (\omega, \eta) \equiv \int\limits_0^1 
\frac{du}{1 + \omega (u - \bar u) - 2 u \bar u \eta + i \epsilon} ,  
\label{eq:J-func}
\end{equation}
whose explicit form and the asymptotic behaviour can be 
found in Appendix~B of Ref.~\cite{Ali:2000ci}. 

\begin{figure}[tb]
\centerline{\psfig{file=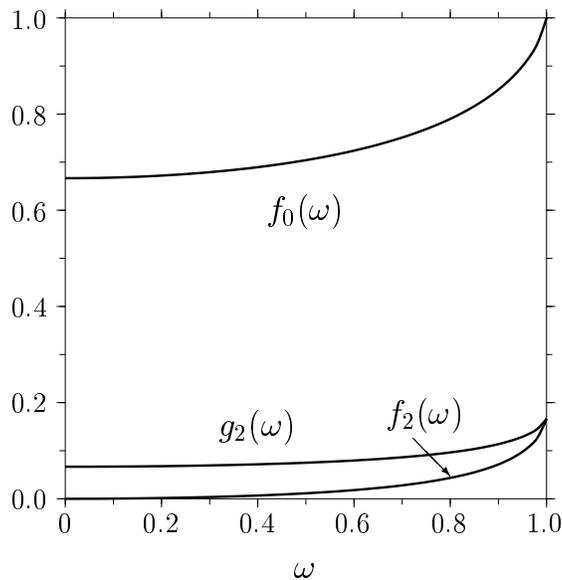,width=.5\textwidth}}
\caption{\label{fig:asymm}%
         The functions $f_0(\omega)$, $f_2(\omega)$, and~$g_2(\omega)$
         describing the large $|q^2|$ asymptotic behaviour of the 
         $\eta^\prime g^* g^*$ effective vertex function, 
         with the virtualities of the gluons having same signs.}
\end{figure}

When the $\eta^\prime$-meson mass is negligible in comparison 
with the total gluon virtuality~$|q^2|$ (i.e., in the limit 
of small~$\eta$), Eq.~(\ref{eq:epgg-res-gen}) reduces to: 
\begin{equation} 
F^{(q)}_{\eta^\prime g^* g^*} (q^2, \omega, 0) =
\frac{4 \pi \alpha_s (|q^2|)}{q^2} \, 
\frac{3 f_{\eta^\prime} \sqrt{N_f}}{N_c}
\left \{ f_0 (\omega) + 6 A_2 (|q^2|) f_2 (\omega) \right \},
\label{eq:epgg-res-zero} 
\end{equation} 
(in accordance with Eq.~(3.10) of Ref.~\cite{Ali:2000ci}) 
as the functions~$G^{(q)}_0 (\omega, \eta)$ 
and~$G^{(q)}_2 (\omega, \eta)$ presented above 
are dominated by the term 
$G^{(q)}_i (\omega, \eta) \simeq \eta \, f_i (\omega)$ 
in this limit, and the functions $f_i (\omega)$ ($i = 0, 2$) 
are defined as follows: 
\begin{eqnarray} 
f_0 (\omega) & = & \frac{1}{\omega^2} 
\left [ 1 - \frac{1 - \omega^2}{2 \omega} 
\ln \frac{1 + \omega}{1 - \omega} \right ] , 
\label{eq:f0} \\ 
f_2 (\omega) & = & \frac{1}{12 \omega^2} 
\left [ 3 (5 - \omega^2) \, f_0 (\omega) - 10 \right ]. 
\label{eq:f2}  
\end{eqnarray} 
The dependence of these functions on the asymmetry 
parameter~$\omega$ is presented in Fig.~\ref{fig:asymm}.
%
%%%%%%%%%%%%%%%%%%%%%%%%%%%%%%%%%%%%%%%%%%%%%%%%%%%%%%%%%%%%%%%%%%%%%%%%%
%
%
%%%%%%%%%%%%%%%%%%%%%%%%%%%%%%%%%%%%%%%%%%%%%%%%%%%%%%%%%%%%%%%%%%%%%%%%%
%
The asymmetry parameter is defined in the 
interval $-1 \le \omega \le 1$, but as the functions~$f_0 (\omega)$ 
and~$f_2 (\omega)$ are symmetric, it is sufficient to present them 
in the region of positive values of the argument. 

If one of the gluons is on the mass shell, for example, 
the second one ($q_2^2 = 0$), the asymmetry parameter 
$\omega = 1$, $\lambda = 1 - \eta$, and it is easy to get 
from Eq.~(\ref{eq:epgg-q0}) the following result:
\begin{equation}
F^{(q)}_{\eta^\prime g^* g} (q_1^2, 0, m_{\eta^\prime}^2) =
\frac{4 \pi \alpha_s (Q^2)}{q_1^2 - m_{\eta^\prime}^2} \,
\frac{f_{\eta^\prime} \sqrt{N_f}}{N_c} 
\int\limits_0^1 \frac{du}{u} \, \phi^{(q)}_{\eta^\prime} (u, Q^2).
\label{eq:epgg-real-g}
\end{equation}
In the approximation of the $\eta^\prime$-meson DAs adopted here, the 
first inverse moment of the quark-antiquark twist-two DA is:
\begin{equation}
\int\limits_0^1 \frac{du}{u} \, 
\phi^{(q)}_{\eta^\prime} (u, Q^2) 
= 3 \left [ 1 + A_2 (Q^2) \right ].
\label{eq:first-inv-moment}
\end{equation}
Thus, in this case, the $\eta^\prime g^* g$ effective vertex function 
is defined by the first inverse moment of the $\eta^\prime$-meson
quark-antiquark DA and it has a pole form, with the pole being at 
$q_1^2 = m_{\eta^\prime}^2$. Of course, this result is not supposed to 
be used in the vicinity of the singular point $q_1^2 \simeq 
m_{\eta^\prime}^2$; the perturbative form sets in at a higher value 
of~$q_1^2$, typically $q_1^2 \simeq 2$~GeV$^2$. Also, as discussed in 
Ref.~\cite{Ali:2000ci}, this pole behaviour results from ignoring the
transverse momentum in the definition of the $\eta^\prime$-meson 
wave-function and, hence, is not physical.

\subsection{The Gluonic Part of the $\eta^\prime g^* g^*$ Vertex}
\label{ssec:epgg-gluon-result}

\begin{figure}[tb]
\centerline{
\psfig{file=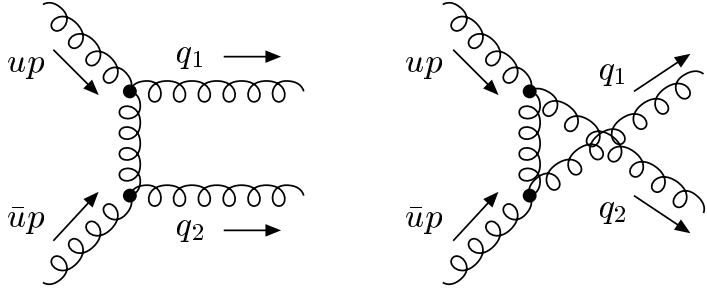,width=.65\textwidth}
}
\caption{\label{fig:diag-gl}% 
         Leading order contribution to the gluonic part 
         of the $\eta^\prime g^* g^*$ vertex.}
\end{figure}

The gluonic part of the $\eta^\prime g^* g^*$ vertex function 
originates from the diagrams presented in Fig.~\ref{fig:diag-gl}. 
%
%%%%%%%%%%%%%%%%%%%%%%%%%%%%%%%%%%%%%%%%%%%%%%%%%%%%%%%%%%%%%%%%%%%%%%%%%
%
%\hspace{90mm}
%
%%%%%%%%%%%%%%%%%%%%%%%%%%%%%%%%%%%%%%%%%%%%%%%%%%%%%%%%%%%%%%%%%%%%%%%%%
% 
To calculate this effective 
vertex function with the help of the invariant 
amplitude~(\ref{eq:MEgg-def}), the corresponding hard-scattering 
kernel~$T^{(g)}_{\rm H}$ can be obtained from the effective four-gluon 
vertex~$V^{A B C D}_{\mu \nu \rho \sigma} (q_1, q_2, q_3, q_4)$, which is 
defined in Eq.~(4.3) of Ref.~\cite{Ali:2000ci}, as follows: 
\begin{equation}
\big [ T^{(g)}_{\rm H} \big ]^{C D}_{\rho \sigma} = 
V^{A B C D}_{\mu \nu \rho \sigma} (- q_1, - q_2, u p, \bar u p) \, 
\varepsilon^{A*}_{1\mu} \varepsilon^{B*}_{2\nu} .  
\label{eq:Tg-def}
\end{equation}
Substituting Eq.~(\ref{eq:glu-PO-real}) into Eq.~(\ref{eq:25}), 
the $\eta^\prime$-meson projection operator onto the two-gluon 
state, the invariant matrix element~(\ref{eq:MEgg-def}) can be 
rewritten in the from:   
\begin{equation} 
{\cal M}^{(g)} = - 
\frac{f_{\eta^\prime}}{4 N_c \sqrt{N_f} \lambda} \int\limits_0^1 du \, 
\frac{\phi^{(g)}_{\eta^\prime} (u, Q^2)}{u \bar u} \,
\frac{\varepsilon^{\rho \sigma \lambda \tau} q_{1\lambda} q_{2\tau}}
     {q_1^2 - q_2^2} \,  \delta_{C D} \, 
\big [ T_{\rm H}^{(g)} \big ]^{C D}_{\rho \sigma} , 
\label{eq:MEgg-real}
\end{equation}
where the parameter~$\lambda$ is defined in Eq.~(\ref{eq:lambda}). 
Comparison of this matrix element with the one given in Eq.~(4.1) of 
Ref.~\cite{Ali:2000ci}, with $C=f_{\eta^\prime}\sqrt{N_f}$, shows that 
the two expressions differ by the factor 
$Q^2/[2 N_f \lambda (q_1^2 - q_2^2)] = Q^2/(2 N_f \lambda \omega q^2)$. 
Note also the difference in the factor~$i$ in the definitions of the 
gluonic projection operators in this paper and in Ref.~\cite{Ali:2000ci} 
(this difference then also reflects itself in the definition of the 
quark-antiquark projection operator). We have now understood  
this mismatch, related to two errors made in Ref.~\cite{Ali:2000ci}: 
First, the factor $1/2$ is due to the identity of gluons in the 
$\eta^\prime$-meson, which was missed in Ref.~\cite{Ali:2000ci}; 
Second, the factor $Q^2/(q_1^2 - q_2^2)$ is required by the Bose 
symmetry of the final gluons in the process $\eta^\prime \to g^* g^*$ 
described by this amplitude, also overlooked in Ref.~\cite{Ali:2000ci}. 
Finally, the parameter~$\lambda$ in the denominator of~${\cal M}^{(g)}$ 
enters as we now take into account the $\eta^\prime$-meson mass; 
$\lambda = 1$ for the case of the massless $\eta^\prime$-meson. 
Taking into account the difference between 
Eq.~(\ref{eq:MEgg-real}) in this paper and Eq.~(4.1) of 
Ref.~\cite{Ali:2000ci} pointed above, the corrected result for the 
gluonic part of the $\eta^\prime g^* g^*$ effective vertex function 
can be obtained from Eqs.~(4.7) and~(4.8) of Ref.~\cite{Ali:2000ci}, 
which now reads as follows: 
\begin{equation}
F^{(g)}_{\eta^\prime g^* g^*} (q^2, \omega, \eta) =
- \frac{4 \pi \alpha_s (Q^2)}{m_{\eta^\prime}^2 \lambda} \, 
\frac{5 f_{\eta^\prime}}{2 \sqrt{N_f}} \, B_2 (Q^2) \, 
{\rm G}^{(g)}_2 (\omega, \eta) , 
\label{eq:GFF-result} 
\end{equation}
where the Gegenbauer moment~$B_2 (Q^2)$ is defined in Eq.~(\ref{eq:DAqg}), 
and the function ${\rm G}^{(g)}_2 (\omega, \eta)$ has the form:
\begin{eqnarray}
{\rm G}^{(g)}_2 (\omega, \eta) & = & 
\frac{2 \eta}{\omega} \int\limits_0^1 du \, 
u \bar u \left ( u - \bar u \right ) 
\frac{\eta + \omega \left ( u - \bar u \right )}
     {1 + \omega \left ( u - \bar u \right ) 
        - 2 u \bar u \eta + i \epsilon}  
\label{eq:Gg1-func} \\
& = & 
\frac{5}{3} + \frac{2}{\eta} - \frac{4 \omega^2}{\eta^2}
+ \frac{1}{2 \omega}
\left [ 1 - \frac{\omega^2}{\eta} \right ]
\left [ 1 - \frac{4 \omega^2}{\eta^2} \right ]
\ln \left | \frac{1 + \omega}{1 - \omega} \right | 
\nonumber \\ 
& + & 
\eta \, \left [ 1 - \frac{2}{\eta} - \frac{2 + \omega^2}{\eta^2}
+ \frac{8 \omega^2}{\eta^3} - \frac{4 \omega^4}{\eta^4}
\right ]
{\rm J} (\omega, \eta) . 
\nonumber
\end{eqnarray}  
This function is symmetric in its first argument under the change 
$\omega \to -\omega$: ${\rm G}^{(g)}_2 (- \omega, \eta) = 
{\rm G}^{(g)}_2 (\omega, \eta)$, in accordance with the 
requirement of the Bose symmetry for the $\eta^\prime g^* g^*$ vertex.

In the limit of the large total gluon virtuality 
($|q^2| \gg m_{\eta^\prime}^2$), the gluonic part of the 
$\eta^\prime g^* g^*$ effective vertex function simplifies 
and can be expressed as follows: 
\begin{equation}
F^{(g)}_{\eta^\prime g^* g^*} (q^2, \omega, 0) =
- \frac{4 \pi \alpha_s (|q^2|)}{q^2} \, 
\frac{5 f_{\eta^\prime}}{2 \sqrt{N_f}} \, B_2 (|q^2|) \, g_2 (\omega) , 
\label{eq:GFF-zero} 
\end{equation}
where the function~$g_2 (\omega)$ has the form: 
\begin{equation} 
g_2 (\omega) = \frac{3 f_0 (\omega) - 2}{6 \omega^2} .  
\label{eq:g2}
\end{equation} 
Here, $f_0 (\omega)$ is the function defined in Eq.~(\ref{eq:f0}). 
Note that the function~$g_2 (\omega)$ is symmetric under the exchange
$\omega \to -\omega$: 
$g_2 (- \omega) = g_2 (\omega)$, in agreement with the observation 
made in Ref.~\cite{Kroll:2002nt}. At $\omega = 0$ it has a non-vanishing 
value: $g_2 (0) = 1/15$, and at the end points, $\omega = \pm 1$, 
it has the value $g_2 (\pm 1) = 1/6$.     
The explicit dependence of~$g_2 (\omega)$ on the asymmetry 
parameter~$\omega$ is presented in Fig.~\ref{fig:asymm}. 

\begin{figure}[tb]
\centerline{
\psfig{file=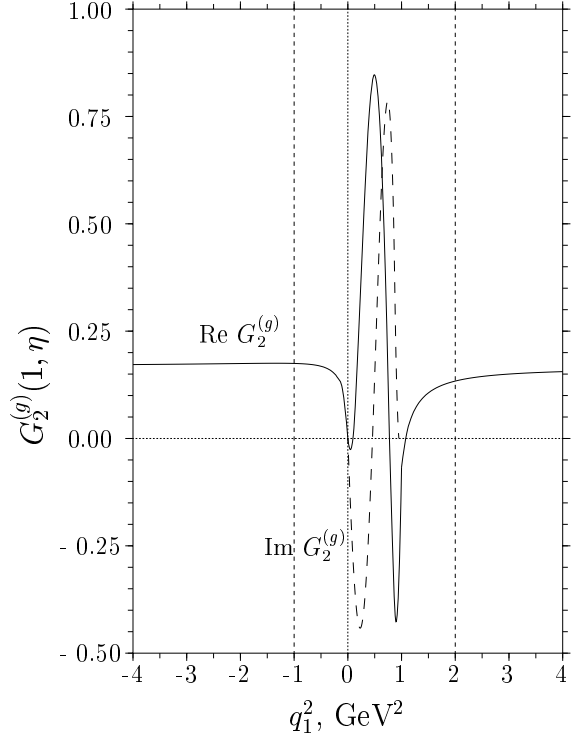,width=.50\textwidth}
}
\caption{\label{fig:Gg2}% 
         The real (solid curve) and imaginary (long-dashed 
         curve) parts of the function~$G^{(g)}_2 (1, \eta)$, where 
         $\eta = m_{\eta^\prime}^2/q_1^2$, as a function of the 
         momentum squared~$q_1^2$ of the virtual gluon. The vertical 
         short-dashed lines, drawn at $q_1^2 = -1~{\rm GeV}^2$  
         and $q_1^2 = 2~{\rm GeV}^2$, demarcate the regions of 
         applicability of the perturbative results, which are to the left 
         and right of these lines for the space-like and time-like gluon 
         virtualities, respectively.}
\end{figure}

If one of the gluons is on the mass shell, say, the second gluon 
($q_2^2 = 0$), the gluonic part of the $\eta^\prime g^* g$ effective 
vertex function can be written in the following form: 
\begin{equation} 
F^{(g)}_{\eta^\prime g^* g} (q_1^2, 0, m_{\eta^\prime}^2) =  
- \frac{4 \pi \alpha_s (Q^2)}{q_1^2 - m_{\eta^\prime}^2} \, 
\frac{5 f_{\eta^\prime}}{2 \sqrt{N_f}} \, B_2 (Q^2) \, 
G^{(g)}_2 (1, \eta) ,   
\label{eq:GFF-mass-shell} 
\end{equation} 
where $\eta = m_{\eta^\prime}^2/q_1^2$, the value 
$\lambda = 1 - \eta$ was taken into account at $\omega = 1$, 
and the function $G^{(g)}_2 (1, \eta)$ is: 
\begin{equation} 
G^{(g)}_2 (1, \eta) =  
\frac{5}{3 \eta} + \frac{2}{\eta^2} - \frac{4}{\eta^3} - 
\frac{1}{\eta} \left [ 1 - \frac{1}{\eta} \right ] 
\left [ 1 - \frac{4}{\eta^2} \right ] \ln (1 - \eta) . 
\label{eq:G2g-end-point} 
\end{equation} 
The dependence of this function on the gluon virtuality~$q_1^2$ 
is presented in Fig.~\ref{fig:Gg2} with the value 
$G^{(g)}_2 (1, 0) = 1/6$ corresponding to the large~$q_1^2$ asymptotics.  
%
%%%%%%%%%%%%%%%%%%%%%%%%%%%%%%%%%%%%%%%%%%%%%%%%%%%%%%%%%%%%%%%%%%%%%%%%%
%
%\hspace{90mm}
%
%%%%%%%%%%%%%%%%%%%%%%%%%%%%%%%%%%%%%%%%%%%%%%%%%%%%%%%%%%%%%%%%%%%%%%%%%
% 
In line with the quark-antiquark part~(\ref{eq:epgg-real-g}) of 
the $\eta^\prime g^* g$ effective vertex function, the gluonic 
part~(\ref{eq:GFF-mass-shell}) has the same pole behaviour at 
$q_1^2 = m_{\eta^\prime}^2$.

To conclude this section, we present the expression 
for the $\eta^\prime g^* g^*$ effective vertex function resulting 
from the perturbative QCD analysis. Combining the 
quark-antiquark~(\ref{eq:epgg-res-gen}) 
and the gluonic~(\ref{eq:GFF-result}) parts of the 
$\eta^\prime g^* g^*$ effective vertex function, 
the perturbative result for the  
vertex can be expressed as follows:
\begin{eqnarray} 
F_{\eta^\prime g^* g^*} (q_1^2, q_2^2, m_{\eta^\prime}^2) & = &  
\frac{4 \pi \alpha_s (Q^2)}{m_{\eta^\prime}^2 \lambda} \, 
\sqrt 3  f_{\eta^\prime}   
\label{eq:VF-total} \\ 
& \times & 
\left [ G^{(q)}_0 (\omega, \eta) 
+ 6 A_2 (Q^2) G^{(q)}_2 (\omega, \eta) - \frac{5}{6} \, 
B_2 (Q^2) G^{(g)}_2 (\omega, \eta) \right ] ,
\nonumber 
\end{eqnarray}
where the total gluon virtuality is $q^2 = q_1^2 + q_2^2$, 
the variables~$\omega$ and~$\eta$ on the right-hand side 
are defined in Eq.~(\ref{eq:parameters}), 
and $Q^2 = |q^2| + m_{\eta^\prime}^2$. Being a perturbative-QCD  
result, this expression is valid in the large~$|q^2|$ region.

As mentioned above, both the quark-antiquark~(\ref{eq:epgg-real-g}) 
and the gluonic~(\ref{eq:GFF-mass-shell}) parts of the 
$\eta^\prime - g$ transition form factor have the phenomenological 
form~(\ref{eq:VF-eta-gamma}) which allows to extract the slowly 
varying function~$H (q_1^2, 0, m_{\eta^\prime}^2)$: 
\begin{equation} 
H (q_1^2, 0, m_{\eta^\prime}^2) = 
\frac{4 \pi \alpha_s (Q^2)}{m_{\eta^\prime}^2} \, 
\sqrt 3 f_{\eta^\prime} \left [ 1 + A_2 (Q^2) -  
\frac{5}{6} \, B_2 (Q^2) G_2^{(g)} (1, \eta) \right ] .  
\label{eq:H-mass-shell}
\end{equation} 
The non-trivial dependence on the $\eta^\prime$-meson 
mass is coming through the function~$G_2^{(g)} (1, \eta)$,  
with $\eta = m_{\eta^\prime}^2/q_1^2$, which is to be traced 
back to the gluonic component of the $\eta^\prime$-meson.

\section{Numerical Analysis} 
\label{sec:numeric} 

As demonstrated earlier, the correction due to the mass of the 
$\eta^\prime$-meson appears already in the leading-twist (twist-two) 
light-cone approximation. It is of interest to know numerically 
the effect of including the $\eta^\prime$-meson mass in the
$\eta^\prime g^* g^*$ effective 
vertex function.
To work this out, we need to specify the input values for the 
various parameters. To that end, we note that the 
$\eta^\prime$-meson decay constant can be related with the 
flavour-singlet decay constant~$f_1 \simeq 1.17 f_\pi$, where 
$f_\pi = 133$~MeV is the $\pi$-meson decay constant, as: 
$f_{\eta^\prime} = f_1 \cos \theta_1$ with the mixing angle 
$\theta_1 \simeq - 9.2^\circ$~\cite{Feldmann:1999uf}. With this, it is  
easy to check that $f_{\eta^\prime} \simeq 2 f_\pi/\sqrt 3$. 
In estimations of the effective vertex function, the strong 
coupling~$\alpha_s (Q^2)$ is used in the two-loop approximation 
with the QCD scale parameter $\Lambda^{(4)}_{\overline{\rm MS}} = 
305$~MeV corresponding to four active flavours ($n_f = 4$).  
The central values of the~$c$- and 
$b$-quark $\overline{\rm MS}$ masses, $\bar m_c = 1.3$~GeV and 
$\bar m_b = 4.3$~GeV~\cite{Hagiwara:fs}, were used for the 
separation of regions with different active-quark flavours  
in the strong coupling~$\alpha_s (Q^2)$. 
In this context we also specify the constrained 
parameters~(\ref{eq:AD-rotate}) and~(\ref{eq:rho's}) 
in the $\eta^\prime$-meson DAs: $\gamma_+^2 \simeq - 0.645$, 
$\gamma_-^2 \simeq -1.421$, $\rho^{(q)}_2 \simeq 2.863$, and 
$\rho^{(g)}_2 \simeq - 0.010$, calculated for $n_f = 4$.  
For the starting point of the evolution scale we take 
$\mu_0^2 = Q_0^2 = 2~{\rm GeV}^2$. We do not include errors on these
parameters, as they are relatively small. 

The largest uncertainty in the $\eta^\prime g^* g^*$ effective 
vertex function is due to the gluonic Gegenbauer 
moment~$B_2 (Q^2)$~(\ref{eq:Bn}). In particular, 
in the approximation~(\ref{eq:DAqg}), 
a fit to the CLEO and L3 data on the 
$\eta^\prime - \gamma$ transition form factor for~$Q^2$ larger 
than 2~GeV$^2$ was recently undertaken in Ref.~\cite{Kroll:2002nt}, 
yielding
\begin{equation} 
A_2 (1~{\rm GeV}^2) = - 0.08 \pm 0.04, 
\qquad 
B_2 (1~{\rm GeV}^2) = 9 \pm 12,  
\label{eq:AB2-fit}
\end{equation} 
where, in the analysis in Ref.~\cite{Kroll:2002nt}, the initial scale in 
the evolution of the Gegenbauer moments was taken as $\mu_0^2 = 1$~GeV$^2$. 
The estimates~(\ref{eq:AB2-fit}) can be translated in terms of the 
universal free parameters~$B^{(q)}_2 (\mu_0^2)$ 
and~$B^{(g)}_2 (\mu_0^2)$ (the Gegenbauer coefficients), fixed at the 
same initial scale~$\mu_0^2$ of the DA evolution, yielding:
\begin{equation} 
B^{(q)}_2 (1~{\rm GeV}^2) = 0.02 \pm 0.17, 
\qquad 
B^{(g)}_2 (1~{\rm GeV}^2) = 9.0 \pm 11.5 .   
\label{eq:B2-fit}
\end{equation} 
We note that this analysis of the $\eta^\prime - \gamma$ transition 
form factor yields an order of magnitude uncertainty in these 
parameters, taking into account the $\pm 1 \sigma$ error. 
In addition, the Gegenbauer moments~$A_2 (1~{\rm GeV}^2)$ 
and~$B_2 (1~{\rm GeV}^2)$ are strongly correlated 
and we refer to Fig.~3 of Ref.~\cite{Kroll:2002nt} 
where the correlation between these quantities is presented. 

The other process which allows to get an independent information  
on the Gegenbauer coefficients in the $\eta^\prime g^* g$ vertex 
is the inclusive decay $\Upsilon(1S) \to \eta^\prime X$.
Recently the $\eta^\prime$-meson energy spectrum was  
measured in this process by the CLEO collaboration and presented in 
Ref.~\cite{Artuso:2002px}. The CLEO data prefers the perturbative-QCD 
motivated form of the $\eta^\prime - g$ transition form 
factor~(\ref{eq:VF-eta-gamma})  for the hard part of the 
$\eta^\prime$-meson energy spectrum, $E_{\eta^\prime} > 0.35 M_\Upsilon$, 
where $M_\Upsilon = 9.46$~GeV is the mass of the $\Upsilon(1S)$-resonance. 
Based on this observation, the form~(\ref{eq:H-mass-shell}) for the 
function $H (q_1^2, 0, m_{\eta^\prime}^2)$ resulting from the 
hard-scattering perturbative-QCD approach for the $\eta^\prime g^* g$ 
effective vertex function was adopted by us in Ref.~\cite{Ali:2003vw}, 
obtaining independent constraints on the Gegenbauer coefficients. 
Unfortunately, the CLEO data on the decay $\Upsilon(1S) \to \eta^\prime X$ 
is statistically very uncertain for the end part of the 
$\eta^\prime$-meson energy spectrum, leaving large uncertainties in the 
determination of the Gegenbauer coefficients from this process alone.  
The combined fit to the data from the $\eta^\prime - \gamma$ transition 
form factor and the $\Upsilon (1S) \to \eta^\prime X$ decay leads 
to more restrictive constraints on these coefficients which 
we have presented and discussed in detail in Ref.~\cite{Ali:2003vw}.
The combined best fit of the Gegenbauer coefficients and the Gegenbauer 
moments yields the following $(\pm~1\sigma)$ values, 
respectively~\cite{Ali:2003vw}:    
\begin{eqnarray}
B_2^{(q)} (\mu_0^2) = - 0.008 \pm 0.054, \qquad
B_2^{(g)} (\mu_0^2) = 4.6 \pm 2.5,
\label{eq:combined-fit} \\
A_2 (\mu_0^2) = - 0.054 \pm 0.029, \qquad
B_2 (\mu_0^2) = 4.6 \pm 2.7, \! \quad
\nonumber  
\end{eqnarray}
where the starting point of the evolution in the $\eta^\prime$-meson 
DAs is taken as $\mu_0^2 = 2$~GeV$^2$.

\begin{figure}[tb]
\centerline{
\psfig{file=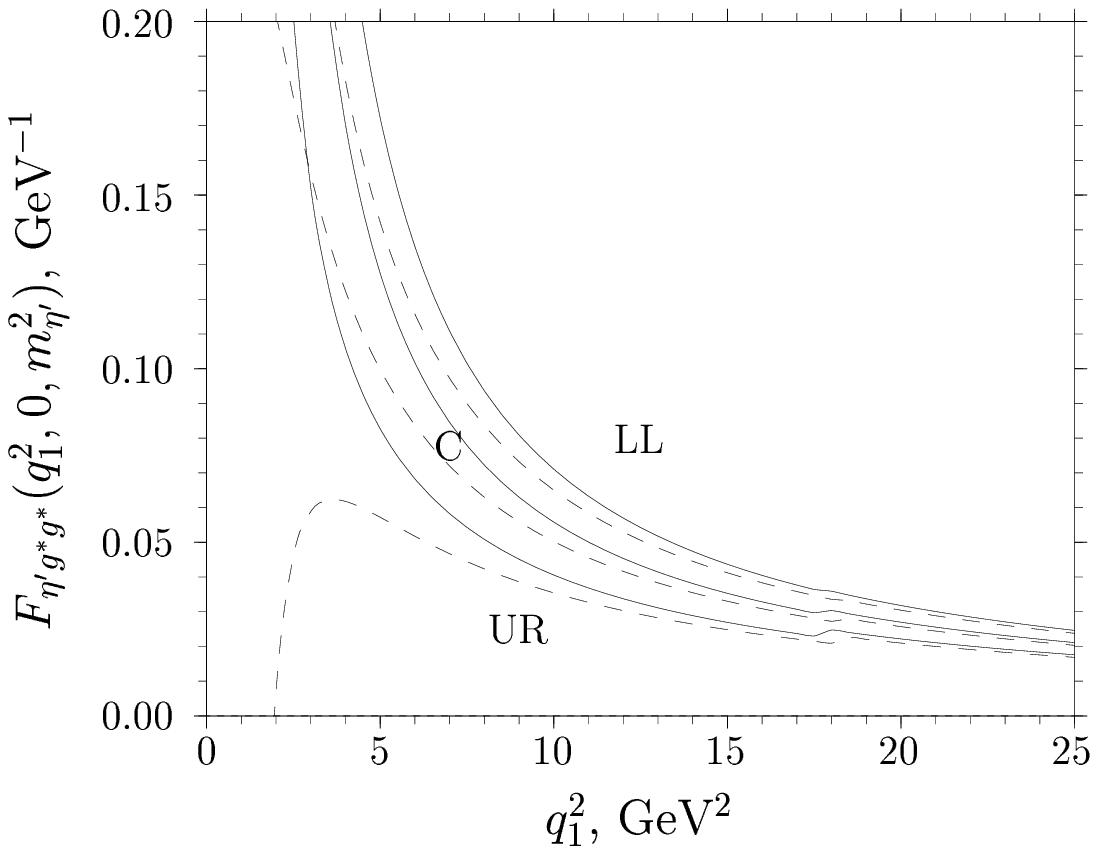,width=.5\textwidth} 
\quad
\psfig{file=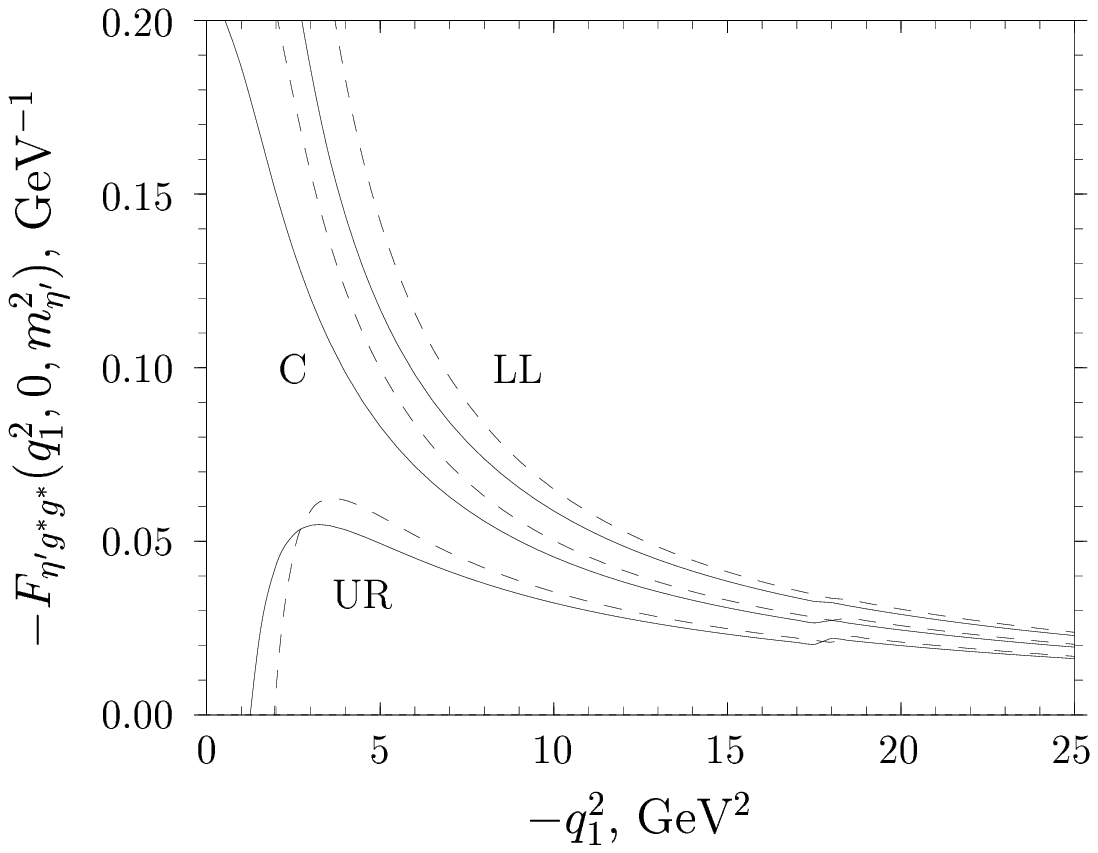,width=.5\textwidth}
}
\caption{\label{fig:FF}% 
         The $\eta^\prime g^* g$ effective vertex function for 
         the time-like (left frame) and space-like (right frame) 
         gluon virtuality~$q_1^2$ when the second gluon is on 
         the mass shell ($q_2^2 = 0$). The solid and dashed curves
         are plotted for the vertex function with and without taking into 
         account the $\eta^\prime$-meson mass, respectively. The labels~C, 
         LL, and~UR correspond to the central, lower-left and upper-right 
         points of the combined best fit of the Gegenbauer coefficients 
         $B_2^{(q)} (\mu_0^2)$ and $B_2^{(g)} (\mu_0^2)$ presented in 
         Fig.~5 of Ref.~\cite{Ali:2003vw} and given in 
         Eq.~(\ref{eq:combined-fit}).} 
\end{figure}

The predictions for the $\eta^\prime g^* g$ effective vertex function
$F_{\eta^\prime g^* g}(q_1^2,0,m_{\eta^\prime}^2)$,  with the second gluon 
on the mass shell ($q_2^2 = 0$), are presented in Fig.~\ref{fig:FF}
for the time-like (left frame)  and space-like (right frame) 
virtuality~$q_1^2$ of the off-shell gluon. 
%
%%%%%%%%%%%%%%%%%%%%%%%%%%%%%%%%%%%%%%%%%%%%%%%%%%%%%%%%%%%%%%%%%%%%%%%%%
%
%\hspace{90mm}
%
%%%%%%%%%%%%%%%%%%%%%%%%%%%%%%%%%%%%%%%%%%%%%%%%%%%%%%%%%%%%%%%%%%%%%%%%%
% 
The labels on the curves~C, LL, and~UR correspond to the central, 
lower-left, and 
upper-right points of the combined best fit of the Gegenbauer 
coefficients presented in 
Fig.~5 of Ref.~\cite{Ali:2003vw}; their values can also be read from  
Eq.~(\ref{eq:combined-fit}).
We note that the effect of including the $\eta^\prime$-meson mass 
in the vertex  function is significant; it becomes crucial for the 
time-like gluon virtuality in the  region $q_1^2 \lesssim 5$~GeV$^2$ for 
the upper-right part of the  combined best-fit region of the Gegenbauer 
coefficients (the curves labeled as UR), but also to a lesser extent for 
the central values of the fit parameters (the curves labeled as C). 
For the space-like gluon virtuality the $\eta^\prime$-meson mass  
effect is numerically not so strong. Nevertheless, it decreases the 
absolute value of the $\eta^\prime g^* g$ transition form factor by 
approximately 10 percent in the region of small gluon virtualities 
($|q_1^2| \lesssim 5$~GeV$^2$) for the Gegenbauer 
coefficients from the lower-left part of the combined best-fit region 
(denoted by the curves labeled as LL) and for the central values of 
the fit (the curves labeled~C).  

Summarizing this section, the results presented in 
Eqs.~(\ref{eq:VF-total}) and 
(\ref{eq:H-mass-shell}) (to be read with Eq.~(\ref{eq:VF-eta-gamma})) are 
our principal analytic results for the vertex functions
$F_{\eta^\prime g^* g^*} (q_1^2, q_2^2, m_{\eta^\prime}^2)$
and $F_{\eta^\prime g^* g} (q_1^2, 0, m_{\eta^\prime}^2)$, respectively.
The vertex function $F_{\eta^\prime g^* g} (q_1^2, 0, m_{\eta^\prime}^2)$ 
is displayed numerically in Fig.~\ref{fig:FF}
in the space-like and time-like regions of the gluon virtuality, 
based on perturbative QCD and our current knowledge of the 
Gegenbauer coefficients. Inclusion of the $\eta^\prime$-meson mass 
effect has led to a rather reliable estimate of this vertex function 
(or the transition form factor) for $q_1^2 \geq 2$~GeV$^2$
for the time-like region, with a power-like fall-off with~$q_1^2$, 
similar to the one seen in the electromagnetic transition form factors 
of the pseudoscalar mesons~\cite{Brodsky:1981rp,Feldmann:1998yc}. 
For the space-like gluon virtuality, the vertex function has a power-like 
fall-off as well, but the current uncertainty in the 
Gegenbauer coefficients still leaves a rather large dispersion in the 
vertex function, in particular for the region 
$\vert q_1^2\vert \lesssim 3$~GeV$^2$. We expect that more precise 
data, such as from the decay $\Upsilon(1S) \to \eta^\prime X$, 
will significantly reduce this parametric uncertainty. In the next 
section, we show that imposing the anomaly constraint on the vertex 
function $F_{\eta^\prime g^* g} (q_1^2, 0, m_{\eta^\prime}^2)$ 
at $q_1^2=0$ considerably reduces this uncertainty for low values 
of~$q_1^2$ in the space-like region.

\section{An Interpolating Formula for the $\eta^\prime g^* g$ 
            Vertex Function for Space-Like Gluon Virtuality} 
\label{sec:interpolation}

As noted earlier, the QCD anomaly determines the value of the vertex 
$F_{\eta^\prime g g} (0, 0, m_{\eta^\prime}^2)$ for on-shell gluons. 
Denoting this by $F^{\rm A}_{\eta^\prime gg}$, 
one has the following expression for this quantity~\cite{Ali:2000ci}:
\begin{equation}
F^{\rm A}_{\eta^\prime gg} = 
- 4 \pi \alpha_s (m_{\eta^\prime}^2) \,
\frac{1}{2 \pi^2 f_{\eta^\prime}}~.
\label{eq:PCAC-value}
\end{equation}
For large off-shellness of the gluons, the form of the vertex function 
is determined by the perturbative QCD and is presented in 
Eq.~(\ref{eq:VF-total}). We would like to write down an expression 
for the vertex function in question which interpolates between 
the non-perturbative result~(\ref{eq:PCAC-value}), applicable at 
$q_1^2 = q_2^2 = 0$, and the perturbative QCD result~(\ref{eq:VF-total}), 
which holds for large virtualities of the gluons. Such a formula is of
considerable phenomenological interest.

Let us concentrate on the case of the  $\eta^\prime g^* g$ effective 
vertex function with one gluon (the second gluon, for definiteness) 
being on the mass shell. Comparing Eqs.~(\ref{eq:epgg-real-g}) 
and~(\ref{eq:GFF-mass-shell}) with the form~(\ref{eq:VF-eta-gamma}), 
it is seen that both the quark-antiquark 
and the gluonic parts of the $\eta^\prime - g$ transition form factor 
have a pole singularity for the time-like virtuality at $q_1^2 = 
m_{\eta^\prime}^2$. However, there is no singularity for the 
space-like region of the gluon virtuality and the vertex function 
$F_{\eta^\prime g^* g} (q_1^2, 0, m_{\eta^\prime}^2)$ is a smooth 
function of~$q_1^2$ in this region (albeit numerically uncertain 
due to the imprecise knowledge of the Gegenbauer coefficients). It is also 
known from 
our earlier work~\cite{Ali:2000ci} that the singularity for the time-like 
region can only be removed by including the transverse momentum effects 
in the $\eta^\prime$-meson wave-function, using the Sudakov resummation 
technique. Since we are ignoring 
the transverse-momentum effects in calculating the vertex function in 
question in this paper, we shall restrict ourselves to the
interpolating function only in 
the space-like region, for which the transverse-momentum effects are 
known to be numerically small~\cite{Ali:2000ci}.

To that end, we work with the function~$H (q_1^2, 0, m_{\eta^\prime}^2)$ 
presented in Eq.~(\ref{eq:H-mass-shell}).
It should be noted that the function~$G_2^{(g)} (1, \eta)$
entering in $H (q_1^2, 0, m_{\eta^\prime}^2)$ is 
very close to its asymptotic value~$1/6$ already at 
$q_1^2 \simeq -1$~GeV$^2$ in the space-like region of the gluon 
virtuality (see Fig.~\ref{fig:Gg2}). 
So, to a very good approximation, the function~$G_2^{(g)} (1, \eta)$
can be replaced by its asymptotic value:
\begin{equation} 
H_{\rm as} (q_1^2) = 
\frac{4 \pi \alpha_s (Q^2)}{m_{\eta^\prime}^2} \, 
\sqrt 3 f_{\eta^\prime} \left [ 1 + A_2 (Q^2) -  
\frac{5}{36} \, B_2 (Q^2) \right ] , 
\label{eq:H-mass-shell-as}
\end{equation} 
with the corresponding vertex function now given by 
$F_{\rm as} (q_1^2) = m_{\eta^\prime}^2 H_{\rm as} (q_1^2)
/(q_1^2 - m_{\eta^\prime}^2)$.
Thus, the dependence of $H_{\rm as}(q_1^2)$ on~$q_1^2$ is coming only 
through~$Q^2 = \vert q_1^2 \vert + m_{\eta^\prime}^2$. 
Defined in this way, the function~$H_{\rm as} (q_1^2)$ is symmetric 
under the change $q_1^2 \to - q_1^2$, and has {\it formally} the    
following limit for on-shell gluons ($q_1^2 = 0$): 
\begin{equation} 
H_{\rm as} (0) = 
\frac{4 \pi \alpha_s (m_{\eta^\prime}^2)}{m_{\eta^\prime}^2} \, 
\sqrt 3 f_{\eta^\prime} 
\left [ 1 + A_2 (m_{\eta^\prime}^2) - 
\frac{5}{36} \, B_2 (m_{\eta^\prime}^2) \right ] .  
\label{eq:H-00}
\end{equation} 
Not unexpectedly, there is a substantial mismatch between the correct 
value of the vertex function for the on-shell gluons 
$F^{\rm A}_{\eta^\prime g g}$, 
%~(\ref{eq:PCAC-value}), 
as determined by the QCD anomaly, and the one obtained 
from the formal limit of the perturbative expression for the 
vertex function $F_{\eta^\prime g^* g} (0, 0, m_{\eta^\prime}^2)= - 
H_{\rm as} (0)$, given in Eq.~(\ref{eq:H-00}). To see this 
quantitatively, we study the numerical consistency of the two expressions, 
which yields the following condition on the Gegenbauer moments 
$A_2 (m_{\eta^\prime}^2)$ and $B_2 (m_{\eta^\prime}^2)$:
\begin{equation}
\frac{2 \sqrt{3} \pi^2  f_{\eta^\prime}^2}{m_{\eta^\prime}^2} \,
\left [ 1 + A_2 (m_{\eta^\prime}^2) -
\frac{5}{36} \, B_2 (m_{\eta^\prime}^2) \right ] =1~.
\label{eq:H-01}
\end{equation}
For the values of the Gegenbauer coefficients given in 
Eq.~(\ref{eq:combined-fit}), this equality is badly violated. 
For example, for the set of the Gegenbauer moments 
$A_2 (m_{\eta^\prime}^2) = - 0.11$ and $B_2 (m_{\eta^\prime}^2) = 2.87$, 
corresponding to the coefficients called LL 
(yielding the largest value for the l.h.s. in Eq.~(\ref{eq:H-01})), 
the l.h.s. in the above equation is about~0.43. For other allowed values 
of the Gegenbauer coefficients, the mismatch is much more pronounced.
This implies the presence of large non-perturbative contributions to 
the moments $A_2 (m_{\eta^\prime}^2)$ and $B_2 (m_{\eta^\prime}^2)$.
However, it is certain that non-perturbative corrections in the 
vertex function $F_{\eta^\prime g^* g}(q_1^2,0,m_{\eta^\prime}^2)$ are to 
be included over a larger region of~$q_1^2$.

To model these non-perturbative effects, we propose the following 
modification of the perturbative 
result~(\ref{eq:H-mass-shell-as}) in the region of the small space-like 
virtualities~$q_1^2$ in terms of the function $\tilde H (q_1^2)$:
\begin{equation}
\tilde H (q_1^2) = H_{\rm as} (q_1^2) + 
\left [ H_A - H_{\rm as} (0) \right ] 
\exp \left [ C_s \, \frac{q_1^2}{m_{\eta^\prime}^2} \right ]~,
\label{eq:H-02}
\end{equation}
with the corresponding vertex function defined as: 
$\tilde F (q_1^2) = m_{\eta^\prime}^2 \tilde H (q_1^2) /
(q_1^2 - m_{\eta^\prime}^2)$.
Here, $H_A = - F^A_{\eta^\prime gg}$ is defined by the anomaly 
value for the $\eta^\prime g g$ vertex in Eq.~(\ref{eq:PCAC-value})  
and~$C_s$ is an arbitrary dimensionless parameter.

This formula gives, by construction, the correct value for the 
$\eta^\prime gg$ vertex  function with the on-shell gluons (i.e., for 
$q_1^2=0)$, reproducing the QCD anomaly, and yields the correct  
perturbative-QCD behaviour for large values of $q_1^2$. Depending on 
the value of the parameter $C_s$, the perturbative result may set in 
rather fast. 
For small values of the gluon virtuality, expanding the exponential 
factor in $q_1^2/m_{\eta^\prime}^2$ gives power corrections.
In the numerical analysis presented below we take $C_s =  2$, as it
allows a smooth interpolation between the anomaly (at $q_1^2=0)$
and the perturbative result for $|q_1^2| > 2$~GeV$^2$.
However, we must admit that this interpolating formula is by no means 
unique, and has to be checked against experimental data on processes 
involving the $\eta^\prime$-meson, or else compared with the results 
obtained using non-perturbative techniques, such as the lattice-QCD. 

%
%%%%%%%%%%%%%%%%%%%%%%%%%%%%%%%%%%%%%%%%%%%%%%%%%%%%%%%%%%%%%%%%%%%%%%%%%
%
%\hspace{90mm}
\begin{figure}[tb]
\centerline{
\psfig{file=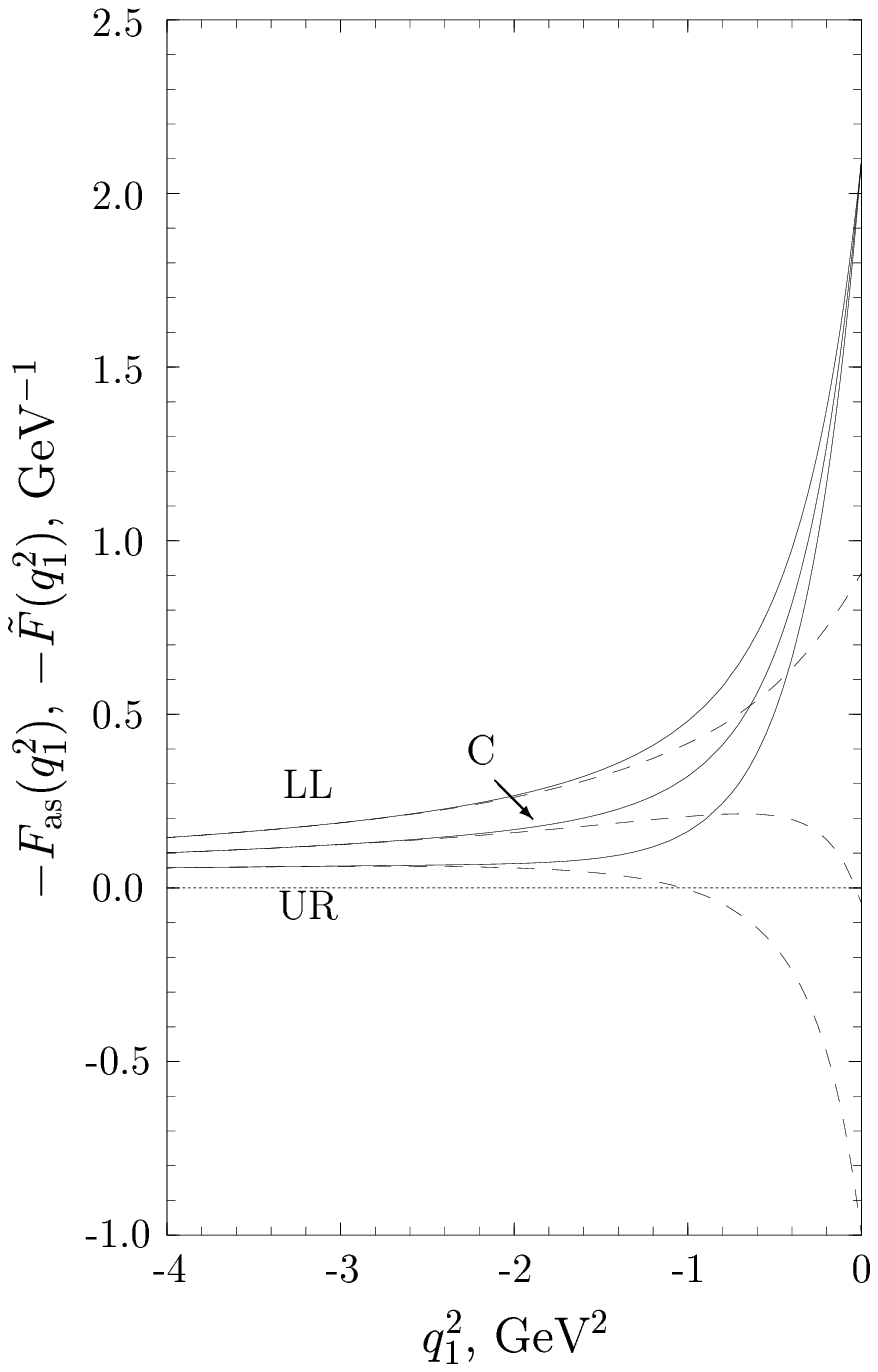,width=.45\textwidth}
\quad
\psfig{file=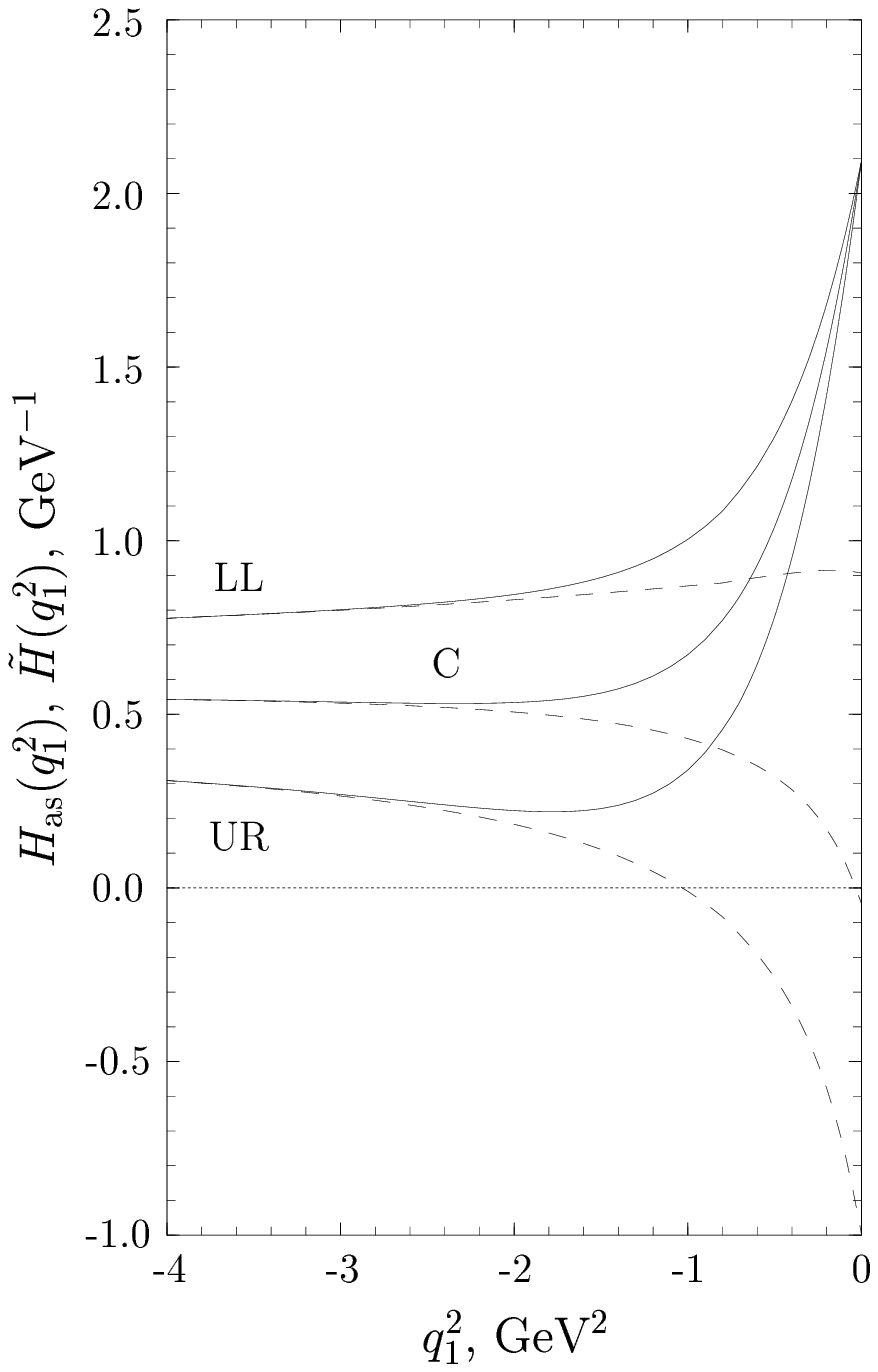,width=.45\textwidth}
}
\caption{\label{fig:asym}% 
         The $\eta^\prime - g$ transition form factor in the perturbative
         QCD approach and using   
         the interpolating formula for the space-like region of the 
         gluon virtuality~$q_1^2$. The left frame shows the functions
         $F_{\rm as}(q_1^2)$ (dashed curves) and $\tilde{F}(q_1^2)$  
         (solid curves). The right frame shows the functions 
         $H_{\rm as} (q_1^2)$ (dashed curves) and $\tilde{H}(q_1^2)$ 
         (solid curves). The labels on the curves are the same as in 
         Fig.~\ref{fig:FF}.}

\end{figure}
%
%%%%%%%%%%%%%%%%%%%%%%%%%%%%%%%%%%%%%%%%%%%%%%%%%%%%%%%%%%%%%%%%%%%%%%%%%
%
The perturbative vertex function $- F_{\rm as}(q_1^2)$ (dashed curves)
and the interpolating function $- \tilde F (q_1^2)$ (solid curves) 
are plotted as functions of~$q_1^2$ in Fig.~\ref{fig:asym} 
(left frame). The function $H_{\rm as} (q_1^2)$ (dashed curves), 
defined in Eq.~(\ref{eq:H-mass-shell-as}), and the 
corresponding interpolating function $\tilde{H}(q_1^2)$ 
(solid curves), introduced in Eq.~(\ref{eq:H-02}), are shown in the 
right frame of this figure.
The curves are marked as C, LL, and UR, corresponding to 
the three sets of values for the Gegenbauer coefficients 
as in  Fig.~\ref{fig:FF}.  
It is obvious that imposing the anomaly condition as a normalization 
point changes the $\eta^\prime g^* g$ vertex function in a  
significant way for low space-like values. For our choice of the 
interpolating function, this change is marked for 
$\vert q_1^2\vert \lesssim 1$~GeV$^2$, increasing the absolute value 
of the function, and reducing the 
theoretical dispersion on the vertex function in this region arising 
due to the imprecise knowledge of the Gegenbauer coefficients in the
perturbative expression.

\section{Conclusions} 
\label{sec:conclusion}  

We have calculated the $\eta^\prime g^* g^{(*)}$ effective vertex function  
in the perturbative QCD approach using the light-cone DAs for the 
$\eta^\prime$-meson with the inclusion of the $\eta^\prime$-meson 
mass. It is shown that if one of the gluons is on the mass shell, 
the pole-like behaviour~(\ref{eq:VF-eta-gamma}) of the $\eta^\prime$~-- 
gluon transition form factor emerges in this approach for 
both the quark-antiquark and the gluonic parts of the form factor, 
and the corresponding function~$H (q_1^2, 0, m_{\eta^\prime}^2)$  
is perturbatively calculated. The Gegenbauer coefficients, 
required for a quantitative analysis of the vertex function, have 
been taken from the combined analysis of the $\eta^\prime - \gamma$ 
transition form factor and the $\Upsilon(1S) \to \eta^\prime X$ decay, 
reported by us earlier in Ref.~\cite{Ali:2003vw}. The corrections due 
to the $\eta^\prime$-meson mass are analyzed numerically, with the  
result that they are important for lower values of the gluon 
virtuality, in particular in the time-like region. An interpolating 
formula connecting
the QCD anomaly value and the perturbative-QCD behaviour of the
$\eta^\prime$~-- gluon transition form factor is presented for the 
space-like gluon virtuality, taking
into account the $\eta^\prime$-meson mass, which modifies the vertex 
function significantly in the region $\vert q_1^2 \vert < 1$ GeV$^2$ and 
reduces the theoretical dispersion in low $\vert q_1^2 \vert$ region 
considerably.

%\newpage    

\section*{Acknowledgements}
%\label{sec:acknowledgements}

We would like to thank Peter Kroll and Kornelija Passek-Kumericki for
helpful correspondence. We also thank Alex Kagan, Christoph Greub, and
Sheldon Stone for numerous discussions. The work of A.Ya.P. has been 
supported by the Schweizerischer Nationalfonds.


\begin{thebibliography}{10}

%\cite{Muta:1999tc}
\bibitem{Muta:1999tc}
T.~Muta and M.~Z.~Yang,
% ``eta' - g* - g transition form factor with gluon content 
% contribution tested,''
Phys.\ Rev.\ D {\bf 61} (2000) 054007
[arXiv:hep-ph/9909484].
%%CITATION = HEP-PH 9909484;%%

%\cite{Ali:2000ci}
\bibitem{Ali:2000ci}
A.~Ali and A.~Y.~Parkhomenko,
%``The eta' g* g* vertex with arbitrary gluon virtualities 
% in the perturbative QCD hard scattering approach,''
Phys.\ Rev.\ D {\bf 65}, 074020 (2002)
[arXiv:hep-ph/0012212].
%%CITATION = HEP-PH 0012212;%%

%\cite{Kroll:2002nt}
\bibitem{Kroll:2002nt}
P.~Kroll and K.~Passek-Kumericki,
%``The two-gluon components of the eta and eta' mesons 
% to leading-twist accuracy,''
Phys.\ Rev.\ D {\bf 67}, 054017 (2003)
[arXiv:hep-ph/0210045].
%%CITATION = HEP-PH 0210045;%%
%\cite{Agaev:2002ek}

\bibitem{Agaev:2002ek}
S.~S.~Agaev and N.~G.~Stefanis,
%``Power corrections to the space-like transition form 
% factor  F(eta' g* g*)(Q**2,omega),''
arXiv:hep-ph/0212318.
%%CITATION = HEP-PH 0212318;%%

%\cite{Kagan:1997qn}
\bibitem{Kagan:1997qn}
A.~L.~Kagan and A.~A.~Petrov,
%``eta' production in B decays: Standard model vs. new physics,''
arXiv:hep-ph/9707354.
%%CITATION = HEP-PH 9707354;%%

%\cite{Atwood:1997bn}
\bibitem{Atwood:1997bn}
D.~Atwood and A.~Soni,
%``B $\to$ eta' + X and the QCD anomaly,''
Phys.\ Lett.\ B {\bf 405}, 150 (1997)
[arXiv:hep-ph/9704357].
%%CITATION = HEP-PH 9704357;%%

%\cite{Kagan:2002dq}
\bibitem{Kagan:2002dq}
A.~L.~Kagan,
%``Beyond the standard model in B decays: Three new results,''
AIP Conf.\ Proc.\  {\bf 618}, 310 (2002)
[arXiv:hep-ph/0201313];
Y.~Chen and A.~L.~Kagan,
Univ. of Cincinnati preprint (in preparation).
%%CITATION = HEP-PH 0201313;%%

%\cite{Artuso:2002px}
\bibitem{Artuso:2002px}
M.~Artuso {\it et al.}  [CLEO Collaboration],
%``Inclusive eta' production from the Upsilon(1S),''
Phys.\ Rev.\ D {\bf 67}, 052003 (2003)
[arXiv:hep-ex/0211029].
%%CITATION = HEP-EX 0211029;%%

%\cite{Ali:2003vw}
\bibitem{Ali:2003vw}
A.~Ali and A.~Y.~Parkhomenko,
%``An Analysis of the Inclusive Decay $\Upsilon (1S) \to \eta^\prime X$ 
% and Constraints on the $\eta^\prime$-Meson Distribution Amplitudes,''
CERN Report CERN-TH/2003-096  
[arXiv:hep-ph/0304278] 
(to appear in Eur. Phys. J. C).
%%CITATION = HEP-PH 0304278;%%

%\cite{Ball:1998je}
\bibitem{Ball:1998je}
P.~Ball,
%``Theoretical update of pseudoscalar meson distribution amplitudes of
% higher twist: The nonsinglet case,''
JHEP {\bf 9901}, 010 (1999)
[arXiv:hep-ph/9812375].
%%CITATION = HEP-PH 9812375;%%

%\cite{Braun:1988qv}
\bibitem{Braun:1988qv}
V.~M.~Braun and I.~E.~Halperin,
%``QCD Sum Rules In Exclusive Kinematics And Pion Wave Function,''
Z.\ Phys.\ C {\bf 44}, 157 (1989)
[Sov.\ J.\ Nucl.\ Phys.\  {\bf 50}, 511 (1989)] 
%YAFIA,50,818-830.1989)].
%%CITATION = ZEPYA,C44,157;%%

%\cite{Braun:1989iv}
\bibitem{Braun:1989iv}
V.~M.~Braun and I.~E.~Halperin,
%``Conformal Invariance And Pion Wave Functions Of Nonleading Twist,''
Z.\ Phys.\ C {\bf 48}, 239 (1990)
[Sov.\ J.\ Nucl.\ Phys.\  {\bf 52}, 126 (1990)]
%\ YAFIA,52,199-213.1990)].
%%CITATION = ZEPYA,C48,239;%%

%\cite{Terentev:qu}
\bibitem{Terentev:qu}
M.~V.~Terentev,
% ``Factorization In Exclusive Processes.
% Form-Factor Of Singlet Mesons In Quantum Chromodynamics,''
Sov.\ J.\ Nucl.\ Phys.\  {\bf 33}, 911 (1981)
[Yad.\ Fiz.\  {\bf 33}, 1692 (1981)].
%%CITATION = SJNCA,33,911;%%

%\cite{Beneke:2002jn}
\bibitem{Beneke:2002jn}
M.~Beneke and M.~Neubert,
%``Flavor-singlet B decay amplitudes in QCD factorization,''
Nucl.\ Phys.\ B {\bf 651}, 225 (2003) 
[arXiv:hep-ph/0210085].
%%CITATION = HEP-PH 0210085;%%

%\cite{Geyer:1999uq}
\bibitem{Geyer:1999uq}
B.~Geyer, M.~Lazar and D.~Robaschik,
%``Decomposition of nonlocal light-cone operators into harmonic 
% operators  of definite twist,''
Nucl.\ Phys.\ B {\bf 559}, 339 (1999)
[arXiv:hep-th/9901090].
%%CITATION = HEP-TH 9901090;%%

%\cite{Geyer:2000ig}
\bibitem{Geyer:2000ig}
B.~Geyer and M.~Lazar,
%``Twist decomposition of nonlocal light-cone operators. 
% II: General  tensors of 2nd rank,''
Nucl.\ Phys.\ B {\bf 581}, 341 (2000)
[arXiv:hep-th/0003080].
%%CITATION = HEP-TH 0003080;%%

%\cite{Radyushkin:1996ru}
\bibitem{Radyushkin:1996ru}
A.~V.~Radyushkin,
%``Asymmetric gluon distributions and hard diffractive 
% electroproduction,''
Phys.\ Lett.\ B {\bf 385}, 333 (1996)
[arXiv:hep-ph/9605431].
%%CITATION = HEP-PH 9605431;%%

%\cite{Ohrndorf:1981uz}
\bibitem{Ohrndorf:1981uz}
T.~Ohrndorf,
%``The Q**2 Dependence Of The Flavor Singlet Pseudoscalar 
% Meson Wave Function In QCD,''
Nucl. Phys. {\bf B186}, 153 (1981).
%%CITATION = NUPHA,B186,153;%%

%\cite{Shifman:1981dk}
\bibitem{Shifman:1981dk}
M.A.~Shifman and M.I.~Vysotsky,
%``Form-Factors Of Heavy Mesons In QCD,''
Nucl. Phys. {\bf B186}, 475 (1981).
%%CITATION = NUPHA,B186,475;%%

%\cite{Baier:1981pm}
\bibitem{Baier:1981pm}
V.N.~Baier and A.G.~Grozin,
%``Meson Wave Functions With Two Gluon States,''
Nucl. Phys. {\bf B192}, 476 (1981).
%%CITATION = NUPHA,B192,476;%%

%\cite{Terentev:wv}
\bibitem{Terentev:wv}
M.~V.~Terentev,
% ``Form-Factor Of Eta-Prime Meson In Quantum Chromodynamics.
% (In Russian),'' 
JETP\ Lett.\ {\bf 33}, 67 (1981) 
[Pisma Zh.\ Eksp.\ Teor.\ Fiz.\  {\bf 33}, 71 (1981)].
%%CITATION = ZFPRA,33,71;%%

%\cite{Belitsky:1998gc}
\bibitem{Belitsky:1998gc}
A.~V.~Belitsky and D.~Muller,
% ``Broken conformal invariance and spectrum of anomalous 
% dimensions in  {QCD},''
Nucl.\ Phys.\ B {\bf 537}, 397 (1999)
[arXiv:hep-ph/9804379].
%%CITATION = HEP-PH 9804379;%%

%\cite{Feldmann:1999uf}
\bibitem{Feldmann:1999uf}
T.~Feldmann,
%``Quark structure of pseudoscalar mesons,''
Int.\ J.\ Mod.\ Phys.\ A {\bf 15}, 159 (2000)
[arXiv:hep-ph/9907491].
%%CITATION = HEP-PH 9907491;%%


%\cite{Hagiwara:fs}
\bibitem{Hagiwara:fs}
K.~Hagiwara {\it et al.}  [Particle Data Group Collaboration],
%``Review Of Particle Physics,''
Phys.\ Rev.\ D {\bf 66}, 010001 (2002).
%%CITATION = PHRVA,D66,010001;%%

%\cite{Brodsky:1981rp}
\bibitem{Brodsky:1981rp}
S.~J.~Brodsky and G.~P.~Lepage,
%``Large Angle Two Photon Exclusive Channels In Quantum 
% Chromodynamics,''
Phys.\ Rev.\ D {\bf 24}, 1808 (1981).
%%CITATION = PHRVA,D24,1808;%%

%\cite{Feldmann:1998yc}
\bibitem{Feldmann:1998yc}
T.~Feldmann and P.~Kroll,
%``Interpolation formulas for the eta gamma and eta' gamma 
% transition  form factors,''
Phys.\ Rev.\ D {\bf 58}, 057501 (1998)
[arXiv:hep-ph/9805294].
%%CITATION = HEP-PH 9805294;%%

\end{thebibliography}
\end{document}